\documentclass[10pt]{article}

\usepackage{amsmath, latexsym, amssymb, amscd, amsfonts, amsthm, epsfig,braket,placeins}

\usepackage[english]{babel}

\DeclareFixedFont{\sfracFont}{U}{euf}{b}{n}{7pt}

\newtheoremstyle{mydefi}
 {15pt}
  {15pt}
  {}
  {}
  {\bfseries}
  {:}
  {.5em}
  {}
\newtheoremstyle{mytheo}
  {15pt}
  {15pt}
  {\slshape}
  {}
  {\bfseries}
  {:}
  {.5em}
  {}

\theoremstyle{mytheo}
\newtheorem{stel}{Theorem}[section]

\theoremstyle{mydefi}

\theoremstyle{mydefi}

\begin{document}

\title{Fisher informations and local asymptotic normality for continuous-time quantum Markov processes}

\author{Catalin Catana$^*$, Luc Bouten$^\dagger$\ , M\u{a}d\u{a}lin Gu\c{t}\u{a}$^*$ \\ \\
\normalsize{$^*$School of Mathematical Sciences, University of Nottingham} \\
\normalsize{University Park, Nottingham NG7 2RD, UK} \\
\normalsize{$^\dagger$ Rijtakker 28, 5768 GT Meijel, The Netherlands}}

\date{}

\maketitle

\vspace{8mm}

\begin{abstract}  
We consider the problem of estimating an arbitrary dynamical parameter of an quantum open system in the input-output formalism. 
For irreducible Markov processes, we show that in the limit of large times the system-output state can be approximated by a quantum Gaussian state whose mean is proportional to the unknown parameter. This approximation holds locally in a neighbourhood of size 
$t^{-1/2}$ in the parameter space, and provides an explicit expression of the asymptotic quantum Fisher information in terms of the Markov generator. 

Furthermore we show that additive statistics of the counting and homodyne measurements also satisfy local asymptotic normality and we compute the corresponding classical Fisher informations. The mathematical theorems are illustrated with the examples of a two-level system and the atom maser.

Our results contribute towards a better understanding of the statistical and probabilistic properties of the output process, with relevance for quantum control engineering, and the theory of non-equilibrium quantum open systems.  
\end{abstract}

\section{Introduction}

The last decades have witnessed rapid progress in the development of quantum technologies \cite{Dowling&Milburn,HarocheRaimond}. These  successes rely on the ability to create and control certain target states which are used as resources for quantum communication \cite{Brassard01}, quantum computing \cite{NC00} or quantum metrology \cite {Giovannetti&Science04}. Effective quantum control is a challenging experimental task, partly because it requires a good understanding of the system's hamiltonian and its interaction with the environment. Therefore, the estimation of dynamical parameters becomes an essential enabling tool for quantum technology. 

\begin{figure}\label{fig:inout}

    \begin{center}  
     \scalebox{0.7}{  \includegraphics[width=0.99\textwidth]{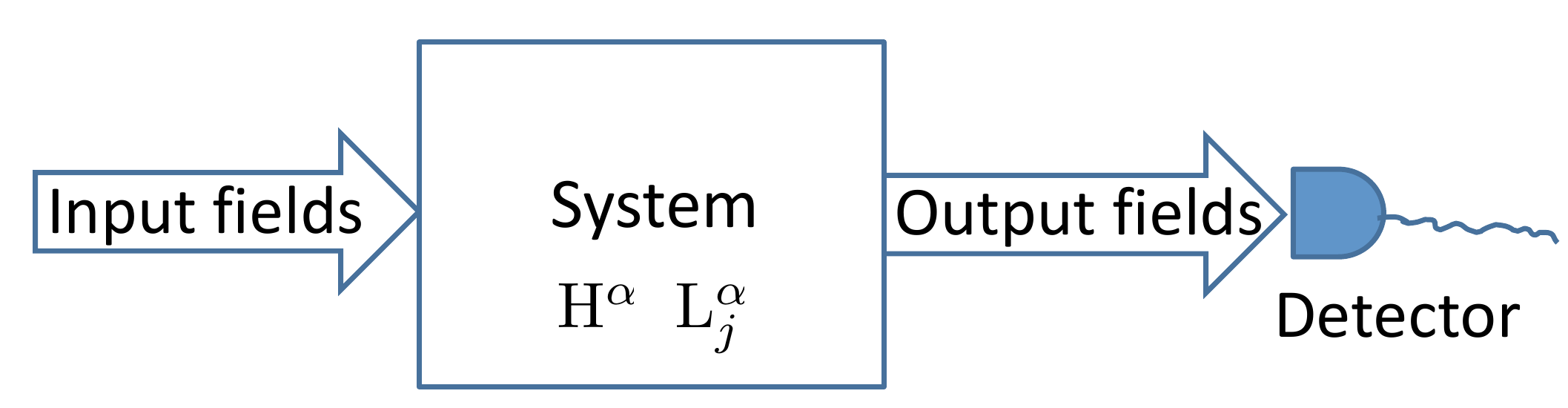}}
    \end{center}
\caption{The input-output formalism. The input fields interact with a system characterized by a Hamiltonian $H^\alpha$ and Lindblad operators $L_j^\alpha$ and evolve into the output fields. The output is continuously monitored and the measurement outcomes are used to infer the unknown parameter $\alpha$.}
 
\end{figure}

In this paper, the system identification problem refers to the estimation of dynamical parameters of an open system in the input-output formalism  \cite{Gardiner&Zoller}, which is routinely used in quantum optics \cite{Car93} and quantum control theory  \cite{Mabuchi&Khaneja}. As illustrated in Figure \ref{fig:inout}, the system is indirectly monitored by performing 
continuous-time measurements in the output channels \cite{Wiseman&Milburn,Breuer02}. The stochastic measurement trajectory is then used for the estimation of an unknown parameter \cite{Mab96,Wiseman&Gambetta,Molmer&Gammelmark}, e.g. the coupling constant between the system and the field. 
Similar problems have been investigated in other system identification scenarios such as quantum channel tomography \cite{Fujiwara}, the estimation of the Hamiltonian of a closed quantum system \cite{Burgarth,Cole}, or the estimation of the Lindblad generator of an 
open system in the Markov approximation \cite{Howard}.

Our study focuses on two distinct aspects of the system identification problem. Firstly we look at the joint system-output state in the limit of large times. We show that this state can be approximated by a quantum Gaussian state whose mean is proportional to the unknown parameter, for a range of parameters localised in a region of the size of the statistical uncertainty $t^{-1/2}$. From a statistical perspective, the quantum statistical model becomes equivalent to a Gaussian one, which allows us to compute the \emph{asymptotic quantum Fisher information}, providing the absolute upper bound on the estimation precision. An alternative computation of the quantum Fisher information can be found in \cite{Molmer&Gammelmark} . 

The second result concerns the  the statistical properties of the \emph{counting and homodyne} continuos-time measurements performed on the output. We show that the \emph{total counts} statistics and the \emph{integrated homdyne currents}, also satisfy local asymptotic normality, in the sense of convergence in distribution to one dimensional Gaussian models with unknown mean and fixed variance. Furthermore we provide explicit expressions for their \emph{classical Fisher informations}. In general such statistics are less informative but computationally much cheaper than standard estimators such as maximum likelihood. It is therefore useful to better understand the statistical power of different output statistics, as it has been shown in recent indepth studies of the atom maser \cite{CatanaGutavanHorssen12,CatanaGutaKypraios}.

Local asymptotic normality for quantum systems has  been  previously investigated for systems of independent qubits \cite{Guta&Kahn, Guta&Janssen08} or independent finite dimensional systems\cite{KahnGuta, Guta&Jencova}. Our work is a generalization for continuous time models of the theory developed for finitely correlated systems in \cite{Guta11,Guta&Kiukas14}. We also point out that the local asymptotic normality of \emph{classical} Markov processes has been derived in \cite{Hopfner88}.

The paper is structured as follows. In the beginning of section \ref{sec:QSM} we introduce the model, an open quantum system whose markovian dynamics depends on an unknown parameter, and the  tools of quantum stochastic calculus needed to prove the main result. Then, using the Trotter-Kato theorem,  we prove a general result concerning the convergence of ergodic  one-parameter semigroups. 
In section \ref{sec:QuantumLAN} we use this convergence theorem  to show that  the joint system-output model converges to a Gaussian model in the limit of large times. We also prove that the Markov processes that describe the continual monitoring of the noise through classical measurements converge to a Gaussian model in the asymptotic regime. 
We illustrate the theoretical results with two examples: a two level system (section \ref{sec:twolevel}) and the atom maser (section \ref{sec:atommsr}).


\section{Background}\label{sec:QSM}
In this section we quickly review the mathematical and physical formalism needed to derive the main result of this paper. We consider a quantum system coupled to an environment through $k$ interaction channels. We assume the environment is memory-less such that the dynamics of the open system is Markovian, i.e. the time evolution of the system is described by the master equation which integrates to a one-parameter semigroup of completely positive operators. The joint dynamics of the system and environment is described by a unitary operator which is the solution of a quantum stochastic differential equation \cite{HuP84} driven by bosonic quantum noises representing the environment degrees of freedom. 
%
The picture is completed by the input-output formalism of Gardiner and Zoller \cite{Gardiner&Zoller} depicted in Fig. \ref{fig:inout}, formalism which describes the evolution of the input fields (initially in the vacuum state) into the output fields; the system can be monitored indirectly through continuous-time measurements in the output (e.g. photon counting or homodyne) to extract information about its state or about the dynamics.  

In the following paragraphs we introduce the formalism of quantum stochastic calculus of Hudson and Parthasarathy \cite{HuP84, Par92}. This is used to derive the equations for the dynamics of the model. The time evolution of the states and operators in the model is given in terms of some one-parameter semigroups. Using the Trotter-Kato theorem we derive a convergence property for these semigroups similar to the results found in \cite{Dav80}. This derivation is essential for proving the final result of this paper, the local asymptotic normality of quantum Markov processes.

%
%
%
%
\subsection{Quantum stochastic calculus }\label{ss:qsc}

Let $\mathcal{H}_s$ be the Hilbert space of the system which we assume to be finite dimensional. The Hilbert space of $k$ independent bosonic fields is  $\mathcal{F}:= \mathcal{F}(L^2(\mathbb{R}^+;\mathbb{C}^k))$, the symmetric  Fock space over the one particle space $\mathbb{C}^k \otimes L^2(\mathbb{R}^{+})\cong L^2(\mathbb{R}^+;\mathbb{C}^k)$. Thus   
$$\mathcal{F}=\mathbb{C}\oplus \bigoplus_{m=1}^{\infty}L^2(\mathbb{R}^+;\mathbb{C}^k)^{\otimes_s m}.$$
 It is useful to define the coherent vector of functions  $f \in L^2(\mathbb{R}^+;\mathbb{C}^k)$ by
\begin{equation}\label{eq:coherent}
e(f)=e^{-\frac{1}{2}||f||^2} \hspace{1mm}\left (1\oplus \bigoplus_{m=1}^{\infty}\frac{f^{\otimes m}}{\sqrt{m!}}\right ).
\end{equation}
The vacuum state is given by $e(0)$ and the inner product of two coherent vector is defined as  $\langle e(f),e(g)\rangle=exp\left \{ -\frac{1}{2}||f||^2 -\frac{1}{2}||g||^2 + \langle f,g\rangle \right \}$. These coherent vectors are linearly independent and their span $\mathcal{D}$ is dense in $\mathcal{F}$. 

Let $f_j, j=1,..,k $ be the  $j$-th component of $f \in  L^2(\mathbb{R}^+;\mathbb{C}^k)$ according to the standard basis in 
$\mathbb{C}^k$. On $\mathcal{D}$ we define the creation process $A_{j,t}^*$, annihilation process $A_{j,t}$ and counting process $\Lambda_{j,t}$ acting in the $j$-th field by
\begin{equation}\label{eq:operators}
\begin{split}
A_{j,t} e(f)&=\langle \chi_{[0,t]}|f_j \rangle e(f)=\int_0^t f_j(s)ds \hspace{1mm}e(f),\\
\langle e(g), A_{j,t}^* e(f)\rangle &=\langle g_j|\chi_{[0,t]} \rangle \langle e(g),e(f) \rangle=\int_0^t \bar{g}_j(s)ds \langle e(g),e(f) \rangle, \\
\langle e(g), \Lambda_{j,t} e(f)\rangle &=\langle g_j|\chi_{[0,t] f_j} \rangle \langle e(g),e(f) \rangle=\int_0^t \bar{g}_j(s)f_j(s)ds \langle e(g),e(f) \rangle.
\end{split}
\end{equation}
For  $0<s<t$ we can write $L^2(\mathbb{R}^+;\mathbb{C}^k)=L^2((0,s);\mathbb{C}^k)\oplus L^2((s,t);\mathbb{C}^k) \oplus L^2((t,\infty);\mathbb{C}^k)$ which combined with the factorization property of the Fock space gives rise to the following tensor product 
$$\mathcal{F}(L^2(\mathbb{R}^+;\mathbb{C}^k))=\mathcal{F}(L^2((0,s);\mathbb{C}^k)) \otimes \mathcal{F}(L^2((s,t);\mathbb{C}^k)) \otimes \mathcal{F}(L^2((t,\infty);\mathbb{C}^k)). $$
   This in turn allows for the identification of the coherent vector with  the product  $e(f) \cong e(f_{s]})\otimes e(f_{[s,t]})\otimes e(f_{[t})$ where $f_{s]}\equiv f\chi_{(0,s)}$, $f_{[s,t]}\equiv f\chi_{[s,t]}$ and $f_{[t}\equiv f\chi_{(t,\infty)}$.

 Let $M^{(k)}_t$ be one of the three processes in \eqref{eq:operators}. Then $M_t^{(k)}$ acts  only on the 'past' and present Fock space i.e. given the factorization property of the Fock space we can write $M_t^{(k)}=M_t^{(k)}\otimes \mathbf{1}_{(t,\infty)}$ and we say $M_t^{(k)}$ is adapted with respect to this factorization. This property is used to define  the stochastic increment   
\begin{equation}
dM^{(k)}_t e(f)\equiv (M_{t+dt}^{(k)}-M_t^{(k)} )e(f)=e(f_{t]})\otimes \left (M_{t+dt}^{(k)}-M_t^{(k)}\right)e(f_{[t,t+dt]}) \otimes e(f_{[t}). 
\end{equation}
Let $X_{1,t}$ and $X_{2,t}$ be two stochastic processes of the type \eqref{eq:operators}, or more generally, processes defined by quantum stochastic differential equations \cite{Par92}
$$
dX_{i,t}= \sum_k a^{(k)}_{i,t} d M^{(k)}_{t}
$$ 
where  $a^{(k)}_{i,t}$ are adapted operator valued coefficients. Then the process $X_{1,t}X_{2,t}$ is adapted and its increment satisfies the \emph{quantum Ito rule}  
\begin{equation}\label{eq:ito}
d(X_{1,t}X_{2,t})=X_{1,t}dX_{2,t}+X_{2,t}dX_{1,t}+dX_{1,t}dX_{2,t}.
\end{equation}

The rules of multiplication of stochastic increments defined at the same time $t$ are given in the following table
\begin{table}[!ht]

 \begin{center}
\begin{tabular}{ l | l  l l  l }
  ~ & $dA^{*}_{i,t}$& $dA_{i,t}$ & $dt$ & $d\Lambda_{i,t}$ \\ \hline
   $dA^{*}_{j,t}$ & 0 & 0 & 0 & 0  \\
   $dA_{j,t}$ & $\delta_{ij}dt$ & 0 & 0 & $\delta_{ij}dA_{j,t}$  \\
   $dt$ & 0 & 0 & 0 & 0 \\
    $d\Lambda_{j}$ &$ \delta_{ij}dA^{*}_{i,t}$ & 0 & 0 & $\delta_{ij}d\Lambda_{i,t}$ \\
\end{tabular}
\end{center}
\end{table}

\subsection{The Markov semigroup of the reduced system evolution}\label{sec.markov.semigroup}
We consider a system with (finite dimensional) Hilbert space $\mathcal{H}_s$ and denote by 
$H$ and $L_j,  j=1,...,k$ the system hamiltonian and coupling with the $k$-bosonic fields representing the environment. 
The joint unitary dynamics  is given by the unique solution  \cite{HuP84} of  the following quantum stochastic differential equation 
\begin{equation}\label{eq:evop}
 dU_t = \left\{\sum_{j=1}^{k}\left(L_j dA_{j,t}^* - L_j^{*}dA_{j,t} -   \frac{1}{2}L_j^{*}L_j dt\right) - iH dt \right\}U_t,
\end{equation}
with $U_0 = \mathbf{1}$ and e.g. $L_j dA_{j,t}^*$ standing for  $L_j \otimes dA_{j,t}^*$.

The joint state of the system and fields at time $t$  is given by
$$
\varrho_t=U_t(\rho_0\otimes \omega)U_t^\dagger,
$$
where $\rho_0$ is  initial state of the system and $\omega= |\Omega\rangle\langle \Omega|$ is the vacuum state of 
the field. 
Using the Ito rules and the fact that the expectation value of the stochastic increments vanishes in the vacuum, one can show that the reduced state  of the system is $\rho_t={\rm Tr}_\mathcal{F}\{\varrho(t)\}$ can be written in terms of a one-parameter semigroup. Indeed by taking time differentials we obtain the following form of the master equation
\begin{equation}\label{eq:master1}
\begin{split}
d\rho_t&=\langle\Omega|\left (     dU_t\rho_0 U_t  + U_t\rho_0 dU_t  + dU_t\rho_0 dU_t         \right )| \Omega\rangle \\
&= \left(-i[H,\rho_t]+\sum_{j=1}^k \left( L_j \rho_t L_j^* - \frac{1}{2}\left\{L_j^* L_j, \rho_t \right\}\right) \right)dt\equiv \mathcal{L}_*(\rho_t) \,dt.
\end{split}
\end{equation}
where $\mathcal{L}_*$ is the Lindblad generator in the Schr\"{o}dinger picture. 
In its integral form, the reduced evolution of the system is given in a terms of a semigroup of trace preserving completely positive maps, characteristic of Markov dynamics
$$
\rho_t=e^{t\mathcal L_*}[\rho_0]\equiv T_{*t} [\rho_0].
$$ 
In the dual, or Heisenberg picture the Lindblad generator is $\mathcal{L}: \mathcal{B}(\mathcal{H}) \to \mathcal{B}(\mathcal{H})$
$$
L(X) =  -i[H,X]+\sum_{j=1}^k \left( L_j^* X L_j - \frac{1}{2} \left\{L_j^* L_j, X \right\}\right)
$$
and the following duality holds for generators as well as for the semigroups 
\begin{equation}\label{eq.duality}
{\rm Tr} (\rho \mathcal{L}(X)) = {\rm Tr} (\mathcal{L}_*(\rho) X), \qquad 
{\rm Tr} (\rho T_{t}(X)) = {\rm Tr} (T_{*t}(\rho) X).
\end{equation}

The Markov dynamics has at least one stationary state, i.e. $T_{*t}[\rho_{ss}]=\rho_{ss}$ or equivalently $\mathcal{L}_*\rho_s=0$. Throughout the paper we will restrict our attention to \emph{irreducible} semigroups which are characterised by the fact that the stationary state is unique and full rank, and any initial state converges in the long run to this stationary state i.e. 
\begin{equation}
\lim_{t\to\infty}T_{*t} (\rho_0) = \rho_{ss}.
\end{equation}
An important property of irreducible semigroups which will be used in the paper is the existence of a spectra gap: the Lindblad generator has a non-degenerate eigenvalue equal to zero (corresponding to the stationary state) and all other eigenvalues have strictly negative real part, cf Theorem 5.4 in \cite{Rivas12}.  



\subsection{Output processes}
We now turn our attention to the evolution of observables, in particular field observable which carry information about the dynamics, and can be measured continuously in time. This is described by the input-output formalism  \cite{Gardiner&Zoller}, in which the  'input' fields are perturbed by the interaction with the system and  propagate out as 'output' fields.

Let $M_{t}$ be one of the fundamental stochastic process of the type \eqref{eq:operators}, which can be seen as an 'input' process $M_t^{in}\equiv M_t $; the corresponding 'output' is obtained by evolving the input with the 
unitary $U_t$: 
\begin{equation}\label{eq:output}
M_t^{out}=U_t^* M_t^{in} U_t.
\end{equation}

The observed stochastic processes  correspond to physical measurements in the environment. We consider two such processes here corresponding to particle counting  and homodyne measurements. 

\subsubsection{Counting measurements}\label{sec.counting}
We first consider the counting process  \cite{Dav76, Car93} obtained by detecting photons in the $i$-th output channel.
The associated quantum stochastic process is $\Lambda^{out}_{i,t}=U_t^* \Lambda_{i,t} U_t$ whose increment is 
\begin{equation}
d\Lambda^{out}_{i,t}=d\Lambda_{i,t}+ L_{i,t}\, dA_{i,t}^* + L_{i,t}^* \, dA_{i,t}+ (L_i^*L_i)_t \, dt,
\end{equation}
where  $X_t=U_t^* (X\otimes\mathbb{I} )U_t$ denotes the evolved system observable $X$. This implies that in the stationary regime the average counts rate per unit of time is $ \langle \Lambda_{i,t} \rangle_{ss}={\rm Tr}(\rho_{ss} L_i^*L_i)$.

For simplicity, in our analysis we will consider the case of a single bosonic field with counting process $\Lambda_t$.
For later use, we introduce a contractions semigroup on $\mathcal{B}(\mathcal{H}_s)$, which can be used to compute the characteristic function of $\Lambda_t$, and therefore encodes the distribution of the counting operators. 
Similarly to the derivation of the master equation, one can show that 
$S_{s,t}:  \mathcal{B}(\mathcal{H}_s) \to \mathcal{B}(\mathcal{H}_s)$ defined by
\begin{equation}\label{eq:defsemigr}
  S^{(s)}_{t}(X) = 
  \left\langle{\Omega}
  \left| U_{t}^{*} \left(X\otimes  e^{ is\Lambda_{t} }\right) U_{t} \right| {\Omega} \right\rangle.
\end{equation}
is a contractions semigroup with generator
$$
\mathcal{L}^{(s)}(X) = \mathcal{L}(X) + (e^{is}-1 ) L^* X L 
$$
In particular, the characteristic function of $\Lambda^{out}_{t}$ for an initial state $\rho_{in}$ is 
$$
\varphi^c_{t}(s) = \mathbb{E}\left(e^{is \Lambda^{out}_{t}}\right) ={\rm Tr} \left( \rho_{in} S_{s,t}(\mathbf{1}) \right).
$$

\subsubsection{Homodyne measurements}
We consider now measurements of a given quadrature of the $i$-th output field.  
Let $Z_t= e^{-i\phi}A_{i,t}+e^{i\phi}A_{i,t}^*$ be the corresponding stochastic process in the environment with $\phi$ defining the measured quadrature. We have that 
\begin{equation}
dZ_t^{out}=e^{-i\phi}dA_{i,t}+e^{i\phi}dA_{i,t}^{*}+e^{-i\phi} L_{i,t}^* dt+e^{i\phi} L_{i,t} dt,
\end{equation}
and therefore 
\begin{equation}
\langle Z_t^{out}\rangle_{ss}=t \,{\rm Tr}(\rho_{ss} ( e^{-i\phi}L_i^{*}+ e^{i\phi}L_i\rangle_{ss})). 
\end{equation}
As in the case of counting, we define the contractions semigroup
$T^{(p)}_{t} : \mathcal{B}(\mathcal{H}_s) \to \mathcal{B}(\mathcal{H}_s)$ 
 \begin{equation}
  T^{(p)}_{t}(X) = \bra{\Omega} U_{t}^{*} \left(X\otimes e^{ ipW_{t}} \right) U_{t}\ket{\Omega}.
  \end{equation}
whose generator is
$$
 \mathcal{L}^{(p)} (X) = \mathcal{L}(X) +  ip(e^{-i\phi} L^{*}X+ Xe^{i\phi}L ) 
- \frac{p^2}{2}X.
$$
Then the characteristic function of $Z_{t}$ for an initial state $\rho_{in}$ is given by
$$
\varphi^Z_{t}(p) := \mathbb{E}\left(e^{ip Z_t}\right) = {\rm Tr} \left( \rho_{in} T^{(p)}_{t}(\mathbf{1}) \right).
$$


\subsection{Convergence of one-parameter semigroups }\label{ss:dyn}


In this section we discuss a general semigroup convergence result which will be used as a technical tool in the local asymptotic normality results. We start with the following Trotter-Kato theorem, cf. \cite{Dav80} (Thm 3.17).
\begin{stel}\label{thm Trotter-Kato}
Let $\mathcal{B}$ be a Banach space and let $\mathcal{B}_0$ be a closed subspace of $\mathcal{B}$. For each $n \ge 0$, let $S_\tau^{(n)}$ be a strongly continuous one-parameter contraction semigroup on $\mathcal{B}$ with generator $\mathcal{L}^{(n)}$. Moreover, let $S_\tau$ be a strongly continuous one-parameter contraction semigroup on $\mathcal{B}_0$ with generator $\mathcal{L}$. Let $\mathcal{D}$ be a core for $\mathcal{L}$. The following conditions are equivalent:
  \begin{enumerate}
\item For all $X \in \mathcal{D}$ there exist  $X^{(n)} \in \mbox{Dom}\left(\mathcal{L}^{(n)}\right)$ such that 
\begin{equation*}
    \lim_{n \to \infty} X^{(n)} = X, \qquad 
    \lim_{n \to \infty} \mathcal{L}^{(n)}\left(X^{(n)}\right) = 
    \mathcal{L}(X).
 \end{equation*}
 \item For all $0 \le s < \infty$ and all $X \in \mathcal{B}_0$ 
    \begin{equation*}
    \lim_{n \to \infty} \sup_{0 \le \tau \le s} 
    \left\|S_\tau^{(n)}(X) - S_\tau(X)\right\| = 0.
    \end{equation*}   
  \end{enumerate}
\end{stel}
We will apply the Trotter-Kato theorem to the following scenario.Let us assume that the generator 
$\mathcal{L}^{(n)}$ can be expanded as  
\begin{equation}\label{eq:qgen}
\mathcal{L}^{(n)}(X)=n\mathcal{L}_0(X)+{\sqrt n} \mathcal{L}_1(X)+\mathcal{L}_2(X)+\mathcal{O}(n^{-1/2}).
\end{equation}
Moreover we assume that $\{\mbox{Ker}(\mathcal{L}_0) + \mbox{Ran}(\mathcal{L}_0)\}$ is dense in $\mathcal{B}(\mathcal{H})$. In this case \cite[Thm.\ 5.1]{Dav80} there exists a projection $P:\ \mathcal{B} \to \mathcal{B}$ 
such that $\mbox{Ker}(P) = \overline{\mbox{Ran}(\mathcal{L}_0)}$ and $\mbox{Ran}(P) = \mbox{Ker}(\mathcal{L}_0)$. With $Q := I-P$ we have $P\mathcal{L}_0P = Q \mathcal{L}_0P = P \mathcal{L}_0Q = 0$, but $Q \mathcal{L}_0 Q \neq 0$. Furthermore, we assume there exists a map $\mathcal{\tilde{L}} :\mathcal{B} \to \mathcal{B}$ such that ${\mathcal{\tilde{L}}}\mathcal{L}_0 = \mathcal{L}_0{\mathcal{\tilde{L}}} = Q$, and that $\mathcal{L}_1(X) \in \mbox{Ran}(\mathcal{L}_0)$ for all $X \in P\mathcal{B}(\mathcal{H})$.
%
\begin{stel}\label{th.semigroup.limit}
Let $S_{\tau}^{(n)}$ be a sequence of  semigroups on a Banach space $\mathcal{B}$ with generators $\mathcal{L}^{(n)} $ satisfying the above assumptions.  Suppose that
\begin{equation*}
  \mathcal{L} := -P\mathcal{L}_1{\mathcal{\tilde{L}}}\mathcal{L}_1+ P\mathcal{L}_2P,
\end{equation*}
generates a one parameter contraction semigroup on $\mathcal{B}_0:= P\mathcal{B}$. Then 
\begin{equation*}
  \lim_{n \to \infty} \sup_{0\le \tau\le T} \Big\|S_{\tau}^{(n)}(X) -
  \exp(\tau\mathcal{L}(X)) \Big\| = 0,
 \end{equation*}  
for all $X \in \mathcal{B}_0$ and $0 \le T < \infty$.  
\end{stel}

\begin{proof} 

For any $X \in \mathcal{B}_0$, we will construct an expansion $X^{(n)}=X+\frac{1}{\sqrt n} X_1 + \frac{1}{n}X_2$. 

Since $\lim_{n \to \infty}X^{(n)}=X$ if we find a suitable choice for $X_1$ and $X_2$ such that $\lim_{n \to \infty} \mathcal{L}^{(n)}\left(X^{(n)}\right) = 
    \mathcal{L}(X)$ then our conclusion follows from the Trotter-Kato theorem. We find that 
$$
\lim_{n \to \infty} \mathcal{L}^{(n)}\left(X^{(n)}\right) = \lim_{n \to \infty} \left(  n\mathcal{L}_0 X+\sqrt n \mathcal{L}_0 X_1+\mathcal{L}_0X_2+\sqrt n \mathcal{L}_1 X+\mathcal{L}_1X_1+\mathcal{L}_2X \right )
$$
Note that $\mathcal{L}_0 X=0$ for  $X \in \mathcal{B}_0$. Moreover, if we choose $X_1=-\mathcal{\tilde{L}}\mathcal{L}_1 X$  then 
$\mathcal{L}_0 X_1+\mathcal{L}_1 X =0$. This leads to 
\begin{align*}
\lim_{n \to \infty} \mathcal{L}^{(n)}\left(X^{(n)}\right) &= \mathcal{L}_0X_2 - \mathcal{L}_1 \mathcal{\tilde{L}}\mathcal{L}_1 X + \mathcal{L}_2 X \\
& =  \mathcal{L}_0X_2 - Q\mathcal{L}_1 \mathcal{\tilde{L}}\mathcal{L}_1 X + Q\mathcal{L}_2 X - P\mathcal{L}_1 \mathcal{\tilde{L}}\mathcal{L}_1 X + P\mathcal{L}_2 X.
\end{align*}
We now choose $ X_2:=  \mathcal{\tilde{L}}Q\mathcal{L}_1 \mathcal{\tilde{L}}\mathcal{L}_1 X - \mathcal{\tilde{L}}Q\mathcal{L}_2 X \in Q\mathcal{B}$ and find 
$$
\lim_{n \to \infty} \mathcal{L}^{(n)}\left(X^{(n)}\right) =- P\mathcal{L}_1 \mathcal{\tilde{L}}\mathcal{L}_1 X + P\mathcal{L}_2 X=\mathcal{L} (X).
$$
The convergence follows from Trotter-Kato theorem. 
\end{proof}
Similar convergence results for the asymptotic behavior of one parameter semigroup with different properties have been derived in \cite{Dav80}.

In this paper we will use a rather special case of Theorem \ref{th.semigroup.limit}. We consider contraction 
semigroups on $\mathcal{B}(\mathcal{H}_s)$, with $\mathcal{H}_s$ a finite dimensional Hilbert space, and such that the first term $\mathcal{L}_0$ in the expansion \eqref{eq:qgen} is the generator of a \emph{irreducible} Markov semigroup. This means that the Schr\"{o}dinger picture generator $\mathcal{L}_{0*}$ has a unique stationary state $\rho_{ss}$ which has full rank, while the Heisenberg picture generator $\mathcal{L}_0$ has $\mathbf{1}$ as the unique zero eigenvector. 
In this case $\mathcal{B}_0 =P\mathcal{B}(\mathcal{H}_s)= \mathbb{C}\mathbf{1}$, where the projection $P$ is defined by $PX = {\rm Tr}(\rho_{ss} X)\mathbf{1}$. 
Moreover, we will show that $\mathcal{L}_0$ leaves $\mathcal{B}_1:= Q\mathcal{B}(\mathcal{H}_s) =\{ X: {\rm Tr}(\rho_{ss}X) =0\}$ invariant and its restriction to this space is invertible. Indeed, if $X\in \mathcal{B}_1$ then 
$$
{\rm Tr}(\rho_{ss}\mathcal{L}_0 (X)) = {\rm Tr}(\mathcal{L}_{0*}(\rho_{ss}) X) =0 
$$
so $\mathcal{L}_0 (X)\in \mathcal{B}_1$. Moreover, let $Y\in \mathcal{B}_1$ be such that $Y$ is orthogonal onto the range 
of $\mathcal{L}_0$ in the sense that 
$
{\rm Tr}(Y \mathcal{L}_0 (X)) =0
$
for all $X$. Then, by using the duality property \eqref{eq.duality} we find ${\rm Tr}(\mathcal{L}_{0*} (Y) X)=0$ for all $X$, which implies that $\mathcal{L}_{0*} (Y) =0$ so that $Y = c\rho_{ss}$. But since $Y\in \mathcal{B}_1$, we have 
${\rm Tr}(\rho_{ss}Y) = 0$ ,which implies $Y=0$. Therefore the range of $\mathcal{L}_0$ is $\mathcal{B}_1$ and the inverse $\tilde{\mathcal{L}}: \mathcal{B}_1\to \mathcal{B}_1$ is well defined.

Besides irreducibility, the only additional condition which will need to be verified when applying Theorem \ref{th.semigroup.limit} is then
\begin{equation}\label{eq.condition.L1}
{\rm Tr}(\rho_{ss} \mathcal{L}_1(\mathbf{1})) = 0.
\end{equation}


\section{Local asymptotic normality for the output state}\label{sec:qlan}

We return now to the Markov model introduced in the previous section, and  assume that the interaction between the quantum system and the environment depends on an unknown parameter $\theta\in \mathbb{R}$. The goal is to find how well we can estimate $\theta$ by when we are allowed to perform arbitrary measurements in the output. 
This question can be approached by invoking 
the quantum Cram\'{e}r-Rao bound \cite{Braunstein&Caves94}, and computing the quantum Fisher information of the output state \cite{Molmer}. However since we are dealing with a time-correlated state, it is not obvious that the quantum Cram\'{e}r-Rao bound is achievable in a `single shot' measurement even in the large time limit. Instead we will take a more fundamental approach aimed at characterising the asymptotic `shape' of the quantum statistical model, which provides both the quantum Fisher information and its asymptotic achievability together with the Gaussian distribution of the optimal estimator. 
The relevant statistical concept is that of local asymptotic normality. We will first briefly review its meaning in the case of quantum statistical models consisting of ensembles of identically prepared systems. After this we formulate the extension to quantum Markov processes, which is one of the main results of the paper. 

\subsection {LAN for ensembles of identically prepared systems}\label{sec:QuantumLAN}

We illustrate the idea of quantum LAN through the simplest example of a one parameter quantum statistical model \cite{Guta11}. Let $|\psi\rangle \in \mathbb{C}^d$ be a pure quantum state and define a family of states 
$$
|\psi_\theta\rangle=e^{-i\theta J}|\psi\rangle
$$ 
indexed by an unknown parameter $\theta \in \mathbb{R}$. The generator $J$ is a self-adjoint operator and we assume that $\langle \psi|J|\psi\rangle=0$. The quantum Cram\'{e}r-Rao bound  asserts that for any measurement and any unbiased estimator $\hat\theta$ (i.e. $\mathbb{E}(\hat\theta)= \theta)$ , the mean square error (MSE) is lower bounded as 
$$
\mathbb{E}\left[(\hat{\theta} - \theta)^2 \right]\geq F_\theta^{-1} 
$$
where $F_\theta$ is the QFI which is determined by the variance of the generator 
$F_\theta = 4 \langle \psi|J^2 |\psi\rangle$. If we are given $n$ identical copies of $|\psi_\theta\rangle$, then the corresponding Fisher information is $nF_\theta$, and therefore $\theta$ can be estimated with error rate scaling as $n^{-1/2}$. 

The philosophy of local asymptotic normality is that for large $n$ the parameter can be localised in a region of size $n^{-1/2+\epsilon}$ with high probability, e.g. by using a proportion $n^{1-\epsilon}$ of the system to produce a rough estimator $\theta_0$. Therefore, in asymptotics it suffices then to understand the local properties of the model, and it is natural to work with the equivalent parametrisation $\theta=\theta_0+\frac{u}{\sqrt n}$ where $\theta_0$ is fixed and known, and $u$ is the `local parameter' to be estimated. Let us denote the joint state of the ensemble by $|\psi_{n,u}\rangle:= |\psi_{\theta_0+u/\sqrt{n}}\rangle^{\otimes n}$, and notice that since we are dealing with pure states, all properties of the statistical model are encoded in the inner products. The following calculation shows that in the limit of large $n$, the local statistical model converges to a limit: 
 \begin{eqnarray}
\lim_{n\to\infty} \langle \psi_{n,u}|\psi_{n,v}\rangle&=&\lim_{n\to\infty}\langle\psi|e^{i(u-v)J/\sqrt n}|\psi\rangle^n=
 \lim_{n\to\infty} \left(1 -\frac{(u-v)^2 F}{8n} + o(n^{-1})\right)^n
\nonumber \\
&=& e^{-(u-v)^2F/8} =\langle\sqrt{F/2}u|\sqrt{F/2}v\rangle.
\label{eq:wkconvex}
\end{eqnarray}
Above, $u,v$ are arbitrary local parameters, $F=F_{\theta_0}$, and the vector $|\sqrt{F/2}u\rangle$ denotes a one parameter model consisting of a coherent state of a one-mode continuous variable system with means $\langle Q\rangle = \sqrt{F/2}u$ and $\langle P\rangle=0$. The convergence \eqref{eq:wkconvex} is an example of local asymptotic normality for pure states quantum models. Its statistical interpretation is that for large $n$, the task of estimating $u$ in the original model becomes equivalent to that of estimating $u$ in the limit Gaussian model. In the case of the latter, measuring $Q/\sqrt{F/2}$ produces an unbiased, normally distributed estimator $\hat{u}$ with mean square error $\mathbb{E}[(\hat{u}-u)^2] = F^{-1}$. The `weak' convergence defined above can be strengthened to an operational notion formulated in terms of quantum channels implementing the convergence \cite{Guta&Kahn,Guta&Kiukas14}, which can be applied to general models with mixed states and arbitrary number of parameters. This provides a rigorous framework for studying asymptotically optimal estimation procedures and establishing the asymptotic normality of the estimator \cite{Gill&Guta}. In this paper we limit ourselves to proving the weak  form of local asymptotic normality 
(in terms of inner products for system and output states) and we refer to \cite{Guta&Kiukas14} on how this can be extended to strong convergence of the output state model.

\subsection { QLAN for the Markov model}

We assume that the Markov dynamics described in section \ref{sec.markov.semigroup} depends on an unknown one-dimensional parameter 
$\theta$, more precisely $H= H_\theta$ and $L_i = L_{i,\theta}$ and the dependence is smooth with respect to $\theta$.  Moreover, we assume that the Markov semigroup is irreducible for any $\theta$. We consider that initially the system is in the pure state $|\chi_0\rangle$, so that the joint initial state of system and environment is $|\Psi(0)\rangle = |\chi_0\rangle \otimes |\Omega\rangle \in \mathcal{H}_s \otimes \mathcal{F}(L^2(\mathbb{R}^+;\mathbb{C}^k))$, where $|\Omega\rangle$ is the joint vacuum state of the bosonic fields.

As in the case of identically prepared systems, we expect that by measuring the (stationary) output for a time $t$, allows us to localise $\theta$ within a neighbourhood of size $t^{-1/2}$.  Therefore, we write $\theta=\theta_0+\frac{u}{\sqrt t}$ with $\theta_0$ fixed and $u\in\mathbb{R}$ the unknown local parameter. 
The evolution of the joint initial state gives rise to the family of pure states  

\begin{equation}
\left | \Psi_t^{u} \right \rangle := U_{t}^{\theta_0+\frac{u}{\sqrt t}} \left |\Psi(0)\right\rangle.
\end{equation}
Since the vector state is only defined up to a complex phase, we make the following choice which allows to establish the convergence of the inner products. Let 

\begin{equation}\label{eq:newstate}
\left | \tilde{\Psi}_t^{u} \right\rangle = e^{i\sqrt{t} u \tilde A}\left | \Psi_t^u \right \rangle, 
\end{equation}
where 
$$\tilde{A} =  {\rm Tr} \left (   \left ( \dot{H} + {\rm Im}  \sum_{i=1}^d \dot{L}_{i} ^{*}L_{i} \right ) \rho_{ss}    \right ) , $$
and $\dot{H}, \dot{L}_{i}$ denote the derivative of $H_\theta$ and $L_{i,\theta}$ with respect to $\theta$, at $\theta=\theta_0$. The following theorem establishes the local asymptotic normality of the joint system and output state.

\begin{stel}\label{qlan}
Consider an open system with space $\mathcal{H}_s$ characterised by its hamiltonian $H_\theta$ and the jump operators $L_{1,\theta}, \dots , L_{k,\theta}$, all of which depend smoothly on an unknown parameter $\theta\in \mathbb{R}$. We assume that the dynamics is irreducible for $\theta= \theta_0$. Let $\theta = \theta_0 + u/\sqrt{t}$ be the local parametrisation around $\theta_0$ and let  $\left| \tilde{\Psi}^u_t \right\rangle$ be the joint system-output state at time $t$, as defined in \eqref{eq:newstate}.

Then the quantum statistical model 
$\left \{\left| \tilde{\Psi}^u_t \right\rangle, u \in \mathbb{R}  \right \}$  converges weakly to the coherent states model $\left\{ \left|\sqrt{\frac{F}{2}}u\right\rangle :u \in \mathbb{R}  \right\}$,  i.e. for $u,v \in \mathbb{R} $
\begin{equation}
\lim_{t \to \infty} \Braket { \tilde{\Psi}^u_t  | { \tilde{\Psi}^v_t }}  = \Braket { \sqrt{\frac{F}{2}}u|\sqrt{\frac{F}{2}}v},
\end{equation}
with limiting quantum Fisher information 
\begin{eqnarray}\label{eq:QFI}
F&=& 8 
\left\langle 
\frac{1}{2}\sum_i \dot{L}_i^{*}\dot{L}_i - {\rm Re}(\dot{H}\tilde B)- {\rm Im}(\sum_i \dot{L}_i^{*}\tilde{B}L_i )
\right\rangle_{ss} \\ 
 \tilde B&=&\tilde{\mathcal{L}}\left(\dot{H}+{\rm Im}\sum_i \dot{L}_i^{*}L_i-
 \left\langle \dot{H}+{\rm Im}\sum_{i=1}^{d}\dot{L}_i^{*}L_i\right\rangle_{ss} \right)\nonumber
\end{eqnarray}

\end{stel}

\begin{proof} 
For a fixed triple $(t,u,v)$ we let $\theta=\theta_0+\frac {u}{\sqrt t}$,  $\theta^\prime=\theta_0+\frac {v}{\sqrt t}$ and define the one parameter contractions semigroup with parameter $\tau$
  \begin{eqnarray*}
  T_{\tau}^{(t,u,v)}&:& \mathcal{B}(\mathcal{H}_s) \to \mathcal{B}(\mathcal{H}_s)\\
  T_{\tau}^{(t,u,v)}&:&X \mapsto e^{-it\tau (\theta-\theta^\prime)\tilde A}\left \langle{\Omega}\left|U_{t\tau}^{\theta_0 +u/ \sqrt{t}*} \left(X\otimes \mathbf{1}\right) U_{t\tau}^{\theta_0 +v/ \sqrt{t}}\right|{\Omega}\right\rangle.
  \end{eqnarray*}
The fact that $T_{\tau}^{(t,u,v)}$ is a semigroup can be shown by differentiation and by using the quantum It\^o rules. Its generator $\mathcal{L}^{(t,u,v)}$ is
 \begin{equation}
  \mathcal{L}^{(t,u,v)}(X) = t\left[
  i (H_{\theta} X -X H_{\theta^\prime}) +
 \sum_{i=1}^{d}\left(L^{*}_{i,\theta} X L_{i,\theta^\prime} -
  \frac{1}{2}\left(L^{*}_{i,\theta} L_{i,\theta} X + X L^{*}_{i,\theta^\prime} L_{i,\theta^\prime} \right) \right)-
  i(\theta-\theta^\prime)\tilde A\right].
  \end{equation}

The inner products are
\begin{equation}\label{eq.inner.products}
\Braket { \tilde{\Psi}_t^u |{ \tilde{\Psi}_t^v }}=e^{-i t (\theta-\theta^\prime)\tilde{A}}
 \langle \chi_0\otimes\Omega
 |U_{t}^{\theta_0 +u/ \sqrt{t} *} U_{t}^{\theta_0 +v/\sqrt{t}} |
 \chi_0 \otimes \Omega\rangle 
 =\left\langle{\chi_0}\left| T_{1}^{(t,u,v)}(\mathbf{1}) \right|{\chi_0}\right\rangle
\end{equation}
By applying Theorem \ref {th.semigroup.limit} we obtain the limit 

\begin{equation}
\lim_{t \to \infty} \bra{\chi_0} T_{1}^{(t,\theta,\theta^\prime)}(\mathbf{1}) \ket{\chi_0}= 
e^{i(v^2-u^2) G}e^{-(u-v)^2 F/8} =e^{i(v^2-u^2)\tau G}{\Braket{\sqrt{\frac{F}{2}}u|\sqrt{\frac{F}{2}}v}},
\end{equation}
where $G\in \mathbb{R}$ is a constant, $\Ket{\sqrt{\frac{F}{2}}v}$ is the one mode 
coherent state with mean $\langle Q\rangle =\sqrt{\frac{F}{2}}v, \langle P\rangle =0$. The details of this calculation 
are found in the appendix \ref{proof.qlan}.

Since the complex phase pre-factor can be absorbed in the definition of the coherent state, we conclude that the system-output model converges weakly to the one parameter coherent state limit model.

\end{proof}

We have shown that asymptotically the joint state of the system and environment are locally statistically equivalent to the Gaussian model of coherent states. The coefficient $F$ in \eqref{eq:QFI} is the quantum Fisher information per unit of time of the  local states \eqref{eq:newstate}.

\section{Local asymptotic normality for measurements on the output}
In this section we prove that additive statistics of continuous measurements on the environment  satisfy (the classical version of) local asymptotic normality. More precisely, let $X_t$ be a real-valued random variable indexed by $t\in \mathbb{R}$ (a summary statistic at time $t$), and that it's distribution depends on an unknown parameter $\theta\in \mathbb{R}$; we suppose that the `amount of information' about $\theta$ grows linearly with $t$. As before, we write $\theta = \theta_0 + u/\sqrt{t}$ and we say that the process satisfies LAN if the following convergence in distribution holds (under $\theta$) as $t\to\infty$
\begin{equation}\label{eq.classical.lan}
\frac{1}{\sqrt{t}} ( X_t - \mathbb{E}_{\theta_0} ( X_t) ) \overset{\mathcal{D}}{\longrightarrow} N(\mu u , \sigma^2).  
\end{equation}
The limit is the normal distribution with mean $\mu u$ and variance $\sigma^2$. Its classical Fisher information is the rescaled limiting Fisher information of $X_t$ and is given by the signal to noise ratio $I= \mu^2/ \sigma^2$. As a consequence of \eqref{eq.classical.lan}, we find that the estimator
$$
\hat{\theta} = \theta_0 + \hat{u}/\sqrt{t} := \theta_0  + (X_t - \mathbb{E}_{\theta_0} ( X_t) )/(t \mu)
$$ 
is asymptotically normal and its mean square error satistisfies
$$
t\mathbb{E}_\theta \left[ (\hat{\theta}  -\theta)^2\right] \longrightarrow I^{-1}. 
$$
To prove \eqref{eq.classical.lan} it suffices to show the convergence of the characteristic functions
\begin{equation}\label{eq:lanm}
\lim_{t\to\infty}\mathbb{E}_{\theta_0+\frac{u}{\sqrt t}}(e^{\frac{is}{\sqrt t}X_t})=e^{iu \mu s - \frac {1}{2}\sigma^2 s^2}.
\end{equation}
Below, we apply this recipe to the total counts and integrated homodyne current statistics.  We stress that these results are for \emph{summary} statistics, i.e. they do not take into account time correlations and typically have smaller Fisher information than the whole stochastic measurement process. More generally, one could consider time averages of more general statistics which depend on the whole detection record over a given time window. A central limit theory for such statistics has been developed in \cite{vanHorssenGuta2} for the case of discrete time quantum Markov chains.

\subsection{Counting process}

We return to the counting process introduced in section \ref{sec.counting} and consider for simplicity that the system is coupled with a single bosonic field. The multi-channel case can be treated similarly.  We assume that the dynamics depends on the unknown one-dimensional parameter $\theta \in \mathbb{R}$, so that $H= H_\theta, L= L_{\theta}$. Recall that $\Lambda_{t}^{out}$ is the counting process resulting from detecting output excitations. We define the counting process  
\begin{equation}\label{eq.rescaled.counting}
Y_{t}= 
\Lambda^{out}_{t}-  t \langle L_{\theta_0}^* L_{\theta_0} \rangle_{ss} 
\end{equation} 
where $ \langle L_{\theta_0}^* L_{\theta_0} \rangle_{ss} $ is a \emph{known} quantity equal to the stationary counting rate when the parameter $\theta$ takes the value $\theta_0$. We will show that the rescaled process $Y_t/\sqrt{t}$ satisfies local asymptotic normality, and can be used to construct an asymptotically normal estimator of $\theta$ whose mean square error can be calculated explicitly.

%
%
%
%
%

\begin{stel}\label{lancounts}
Consider an open system with space $\mathcal{H}_s$ characterised by its hamiltonian $H_\theta$ and a jump operator $L_{\theta}$, both of which depend smoothly on an unknown parameter $\theta\in \mathbb{R}$. We assume that the dynamics is irreducible for $\theta= \theta_0$. Let $\theta = \theta_0 + u/\sqrt{t}$ be the local parametrisation around $\theta_0$ and let  $Y_t$ be the counting process defined in \eqref{eq.rescaled.counting}. Then $Y_t$ satisfies local asymptotic normality, i.e.  the following convergence in distribution holds under $\theta = \theta_0 + u/\sqrt{t}$
\begin{equation}\label{lan.counting}
\frac{1} {\sqrt{t}}Y_t \overset{\mathcal{D}} {\longrightarrow}N(\mu_c u , V_c ).
\end{equation}
The limit is the normal distribution with mean $\mu_c u$ and variance $V_c$, both of which can be computed explicitly (see end of proof). In particular, the asymptotic rescaled classical Fisher information of $Y_t$ is given by 
$$
I_c = \frac{\mu_c^2}{V_c} \leq F
$$
and the estimator $\hat{\theta}_t:= \theta_0 + Y_t/ ( t\mu_c)$ is asymptotically normal and satisfies 
$$
\lim_{t\to \infty}  t \mathbb{E} \left[( \hat{\theta}_t- \theta)^2\right] = I_c^{-1}.
$$

\end{stel}

\begin{proof} 
To prove  \eqref{lan.counting} it suffices to prove the convergence of the corresponding characteristic functions
\begin{equation}\label{eq.characteristic.fct}
\lim_{t \to \infty}\mathbb{E}\left(e^{ isY_{t}/\sqrt{t} }  \right) = e^ {i u \mu_c s - \frac{1}{2} V_c s^2 }.
\end{equation}
To establish this, we introduce a family of contractions semigroups 
$S_\tau^{(t,u,s)}: \mathcal{B}(\mathcal{H}_s)\to \mathcal{B}(\mathcal{H}_s)$, where $u,s$ are considered fixed and 
$t$ is an index  playing the role of $n$ in Theorem \ref{th.semigroup.limit}. The semigroups are given by 
\begin{equation}\label{eq.semigroup.rescaled.counting}
S_\tau^{(t,u,s)}(X) = 
\left\langle \Omega \left| U_{t\tau}^{\theta_0 + u/\sqrt{t}*} \left(X\otimes  e^{is Y_{t\tau}/\sqrt{t}} \right)U_{t \tau}^{\theta_0 + u/\sqrt{t}*} \right|\Omega \right\rangle.
\end{equation}
Using Theorem \ref{th.semigroup.limit}  we will show that 
\begin{equation}\label{eq.TK.counting}
\lim_{t \to \infty} S^{(t,u,s)}_{\tau}(\mathbf{1}) = 
e^{ius\tau \mu_c- \frac{s^2}{2}\tau V_c} \mathbf{1},
\end{equation}
where $\mu_c$ and $V_c$ are constants whose explicit expression is given at the end of the proof.

The limit \eqref{eq.characteristic.fct} follows from \eqref{eq.TK.counting} by setting $\tau=1$ and taking expectation 
with respect to the system's initial state on both sides. 
The asymptotic rescaled Fisher information of $Y_t$ is the Fisher information of the Gaussian shift model 
$\{N(\mu_c u , V_c): u\in \mathbb{R}\}$ which is equal to $I_c = \mu_c^2/V_c$.

The proof of the limit \eqref{eq.TK.counting} can be found in Appendix \ref{app.counting}.

\end{proof}

%
%

\subsection{Homodyne measurement}\label{sec.homodye}

In the same setup as the previous section, we consider the homodyne measurement with quadrature angle $\phi$, described by the quantum output process (integrated homodyne current)
$Z_t= e^{-i\phi}A^{out}_k(t)+e^{i\phi}A_k^{out*}(t)$. As before, we define the random variable which is centred at $\theta=\theta_0$
\begin{equation}\label{eq.wt}
W_{t}=Z_{t}- t \langle e^{-i\phi}L_{\theta}^{*}+ e^{i\phi}L_{\theta}\rangle_{ss}.
\end{equation}
We will show that $W_t$ satisfies local asymptotic normality, as $t\to\infty$.

\begin{stel}\label{lanhom}
Consider an open system with space $\mathcal{H}_s$ characterised by its hamiltonian $H_\theta$ and a jump operator $L_{\theta}$, both of which depend smoothly on an unknown parameter $\theta\in \mathbb{R}$. We assume that the dynamics is irreducible for $\theta= \theta_0$. Let $\theta = \theta_0 + u/\sqrt{t}$ be the local parametrisation around $\theta_0$ and let  $W_t$ be the integrated homodyne current defined in \eqref{eq.wt}. Then $W_t$ satisfies local asymptotic normality, i.e.  the following convergence in distribution holds under $\theta = \theta_0 + u/\sqrt{t}$
\begin{equation}\label{lan.homo}
\frac{1} {\sqrt{t}}W_t \overset{\mathcal{D}} {\longrightarrow}N(\mu_h u , V_h ).
\end{equation}
The limit is the normal distribution with mean $\mu_c u$ and variance $V_c$, both of which can be computed explicitly (see end of proof).
 In particular, the asymptotic rescaled classical Fisher information of $W_t$ is given by 
$$
I_h = \frac{\mu_h^2}{V_h} \leq F
$$
and the estimator $\hat{\theta}_t:= \theta_0 + W_t/ ( t\mu_h)$ is asymptotically normal and satisfies 
$$
\lim_{t\to \infty}  t \mathbb{E} \left[( \hat{\theta}_t- \theta)^2\right] = I_h^{-1}.
$$

\end{stel}

\begin{proof} 

To prove \eqref{lan.counting} it suffices to show the convergence of characteristic functions 
\begin{equation}\label{char.fct.homo}
\lim_{t \to \infty}\mathbb{E}\left(e^{\frac{ip}{\sqrt t}W_{t}}\right) = e^ {i up \mu_h - \frac{1}{2} p^2V_h }.
\end{equation}


We define a family of one-parameter contractions semigroups $T^{(t,u,p)}_\tau:\mathcal{B}(\mathcal{H}_s)\to\mathcal{B}(\mathcal{H}_s)$  indexed by $(t,u,p)$, with $u,p$ fixed and $t$ playing the role of $n$ in Theorem \ref{th.semigroup.limit}. 
The semigroups are given by
  \begin{equation}
  T_{\tau}^{(t,u,p)}(X) = 
\left\langle \Omega\left|
U_{t\tau}^{\theta_0+u/\sqrt{t}*} \left(X\otimes e^{ip W_{t\tau}/\sqrt{t}} \right) U_{t\tau}^{\theta_0+u/\sqrt{t}} \right|\Omega\right\rangle,
  \end{equation}
Using Theorem \ref{th.semigroup.limit}  we will show that 
\begin{equation}\label{eq.TK.homo}
\lim_{t \to \infty} T^{(t,u,p)}_{\tau}(\mathbf{1}) = 
e^{iup\tau \mu_h- \frac{p^2}{2}\tau V_h} \mathbf{1},
\end{equation}
where $\mu_h$ and $V_h$ are constants whose explicit expression is given at the end of the proof.

The limit \eqref{char.fct.homo} follows from \eqref{eq.TK.homo} by setting $\tau=1$ and taking expectation 
with respect to the system's initial state on both sides. 
The asymptotic rescaled Fisher information of $W_t$ is the Fisher information of the Gaussian shift model 
$\{N(\mu_h u , V_h): u\in \mathbb{R}\}$ which is equal to $I_h = \mu_h^2/V_h$.

The proof of the limit \eqref{eq.TK.homo} can be found in Appendix \ref{app.homo}.

%
%

\end{proof}

\section{Examples}

In this section we apply the general results to two examples, a two level system and the atom maser. 

\subsection{Two-level system}\label{sec:twolevel}

Let us consider a two level atom with Hilbert space  $\mathcal{H}_s = \mathbb{C}^2$ interacting with an electromagnetic field with jump operator  $L_\theta = \theta\sigma_- + z\mathbf{1}$, where $ z \in \mathbb{C}$ and $\theta \in \mathbb{R}$ is an unknown parameter. We choose the interaction Hamiltonian  $H_\theta=\frac{i}{2}\theta(\bar{z}\sigma_- - z\sigma_+)$ where $\sigma_\pm = \frac{1}{2}(\sigma_x \pm i\sigma_y)$. The reduced dynamics of the atom has a stationary state 
\begin{equation}
\rho_{ss}(\theta) = \begin{pmatrix}
 a  & b \\
 c       & 1-a\\
  \end{pmatrix},\qquad
  a=\frac{4|z|^2}{8|z|^2+\theta^2}, \quad b=-\frac{\theta z}{2 |z|^2}a , \quad c=-\frac{\theta \bar{z}}{2 |z|^2}a.
\end{equation}
Since the identity operator  spans the kernel of $\mathcal{L}_0$ the projection $P$ can be defined by $PX={\rm Tr}(\rho_{ss}X)$. 
Let us define the following vectors in $\mathbb{C}^2$ 
  \begin{equation}\label{eq.basis}
  e_1 = \sigma_z + \tilde{a}\mathbf{1},\qquad e_2 = \sigma_+ + \tilde{b}\mathbf{1},\qquad 
  e_3 = \sigma_-+ \tilde{c} \mathbf{1},\qquad e_4 = \mathbf{1},
  \end{equation}  
where $\tilde{a} =1-2a ,\tilde{b}= -c,\tilde{c}=-b$ are chosen in such a way that they span the orthogonal complement of $P$ and  the assumptions of section \ref{ss:dyn} hold.  

Our goal is to compute the quantum Fisher information of the output, and the classical Fisher informations for counting 
and homodyne, as described in Theorems \ref{qlan}, \ref{lancounts} and respectively \ref{lanhom}. For clarity, we present here some of the main ideas, while the details of the computations can be found in Appendix \ref{app.2levels}.
\subsubsection{LAN for output  states}
Following the method of Section \ref{sec:qlan}, we localise the unknown parameter as $\theta=\theta_0+\frac{u}{\sqrt t}$, and we expand the generator \eqref{eq.generator.qlan}  with respect to $\sqrt{t}$ and find the first three terms
\begin{equation}
\begin{split}
  &\mathcal{L}_0(X) =   i[H,X] +\theta_0^2  \left(  \sigma_+ X \sigma_- - \frac{1}{2}\{\sigma_+ \sigma_-,X \}\right),\\
  &\mathcal{L}_1(X) =   \frac{i}{\theta_0}(vH X - uXH )+ \theta_0 
 \left( (v+ u) \sigma_+ X \sigma_- - \left(v\sigma_+ \sigma_- X+uX\sigma_+\sigma_-\right ) \right), \\
  &\mathcal{L}_2(X) =
   uv  \sigma_+ X \sigma_- -
\frac{1}{2} \left (  {v^2} \sigma_+\sigma_-X +
{u^2}X \sigma_+\sigma_- \right ). 
\label{eq.els}
  \end{split}\end{equation}
Following the same steps as in the proof of Theorem \ref{qlan}, one can show that the quantum Fisher information is equal to 
\begin{equation}
F(\theta_0)=\frac{128 |z|^4}{(8|z|^2 + \theta_0^2)\theta_0^2}. 
\end{equation}

We note that the QFI depends only on the ratio  $|z|/\theta_0$ and diverges when the coupling constant vanishes.

\subsubsection{Classical measurements}
If we consider  the counting process we remark that the average number of radiated photons 
\begin{equation}
\begin{split}
\langle L_{\theta_0}^{*}L_{\theta_0}\rangle_{ss} &=\frac{\theta_0^2}{2}\langle\sigma_z \rangle_{ss} +\theta_0\langle \bar{z}\sigma_-+z\sigma_+\rangle_{ss} +\frac{\theta_0^2}{2}  + |z|^2 \\
&=-\frac{\theta_0^2}{2}\tilde{a}-\theta_0(z\tilde{b}+\bar{z}\tilde{c})+\frac{\theta_0^2}{2}  + |z|^2 = |z|^2,
\end{split}
\end{equation}
equals  the intensity of the driving laser (as expected) and therefore the rescaled asymptotic classical Fisher information of the total counts statistics is zero.

For  homodyne measurements the generator of dynamics  can be expanded in the usual perturbation series with coefficients
\begin{eqnarray*}
\mathcal{L}_0(X) &= &  \frac{\theta_0}{2}[\bar{z}\sigma_--z\sigma_+,X] +\theta_0^2  \left(  \sigma_+ X \sigma_- - \frac{1}{2}\{\sigma_+ \sigma_-,X \}\right),\\
\mathcal{L}_1(X) &=&  iu[\dot{H},X]+   u ( \dot{L}^{*}XL+ L^{*}X\dot{L})\\
 && -\frac{u}{2}\left( (  \dot{L}^{*}L+ L^{*}\dot{L}  )X+X( \dot{L}^{*}L+ L^{*}\dot{L} )\right) \\
&& + ip(e^{-i\phi}L^{*}X+ Xe^{i\phi}L) - ip\langle e^{-i\phi}L^{*}+ e^{i\phi}L\rangle_{ss} X \\
\mathcal{L}_2(X) &= & i up(e^{-i\phi}\dot{L}^{*}+ e^{i\phi}\dot{L})X 
+{u^2} \dot{L}^{*}X\dot{L} 
 -\frac{u^2}{2} (\dot{L}^{*}\dot{L} X+X \dot{L}^{*}L)- \frac{p^2}{2}X. 
\end{eqnarray*}
 The numerator in the Fisher information   is defined in terms of the mean value
\begin{equation}
\begin{split}
\langle e^{-i\phi} L^{*}+e^{i\phi}L\rangle_{ss} &=\theta_0\langle e^{-i\phi}\sigma_+ +e^{i\phi}\sigma_- \rangle_{ss} + 2{\rm Re}(ze^{i\phi}) \\
&=-\theta_0(e^{-i\phi}\tilde{b}+e^{i\phi}\tilde{c})+2 {\rm Re}(ze^{i\phi})\\
&= 2{\rm Re}(ze^{i\phi})-4{\rm Re}(ze^{i\phi})\tilde{a}
\end{split}
\end{equation}
The denominator is the coefficient of $-p^2/2$ in the expression $-P\mathcal{L}_1\tilde{\mathcal{L}}\mathcal{L}_1(I)+P\mathcal{L}_2(I)$. Remark that the coefficient of $p^2/2$ in $P\mathcal{L}_2(I)$ is simply $-1$.

Starting from 
\begin{equation}
\begin{split}
\mathcal{L}_1(\mathbf{1})&=ip(e^{-i\phi}L^{*}+e^{i\phi} L )-\langle e^{-i\phi} L^{*}+e^{i\phi} L\rangle_{ss}\\
&=ip\theta_0(e^{-i\phi}e_2+e^{i\phi}e_3).
\end{split}
\end{equation}
We find that
\begin{eqnarray*}
\tilde{\mathcal{L}}\mathcal{L}_1(I)&=&\frac{ip}{\theta_0(\theta_0^2+8|z|^2)}\left(   -4{\rm Re}((e^{i\phi}z)\theta_0 e_1+ \left((-2\theta_0^2-8|z|^2)e^{-i\phi}+8z^2 e^{i\phi}\right)e_2 \right.\\
&&+\left.\left((-2\theta_0^2-8|z|^2)e^{i\phi}+8\bar{z}^2 e^{-i\phi}\right)e_3    \right).
\end{eqnarray*}
Then the denominator in the Fisher information reads
\begin{equation}
B_h=1+\frac{2}{(\theta_0^2+8|z|^2)^3}\left(  \theta_0^4 \left(4 {\rm Im}^2(e^{i\phi}z)-16|z|^2\right)+192\theta_0^2|z|^4+512|z|^4{\rm Im}^2(e^{i\phi}z)  \right)
\end{equation}
And the Fisher information is given by
\begin{equation}
I_{h}=\frac{A_h^2}{B_h},
\end{equation}
with
\begin{equation}
A_h=\frac{64\theta_0|z|^2 Re(e^{i\phi}z)}{(\theta_0^2+8|z|^2)^2}.
\end{equation}


\begin{figure}\label{fig:homd}
  \hfill
  \begin{minipage}[t]{.45\textwidth}
    \begin{center}  
       \includegraphics[width=0.99\textwidth]{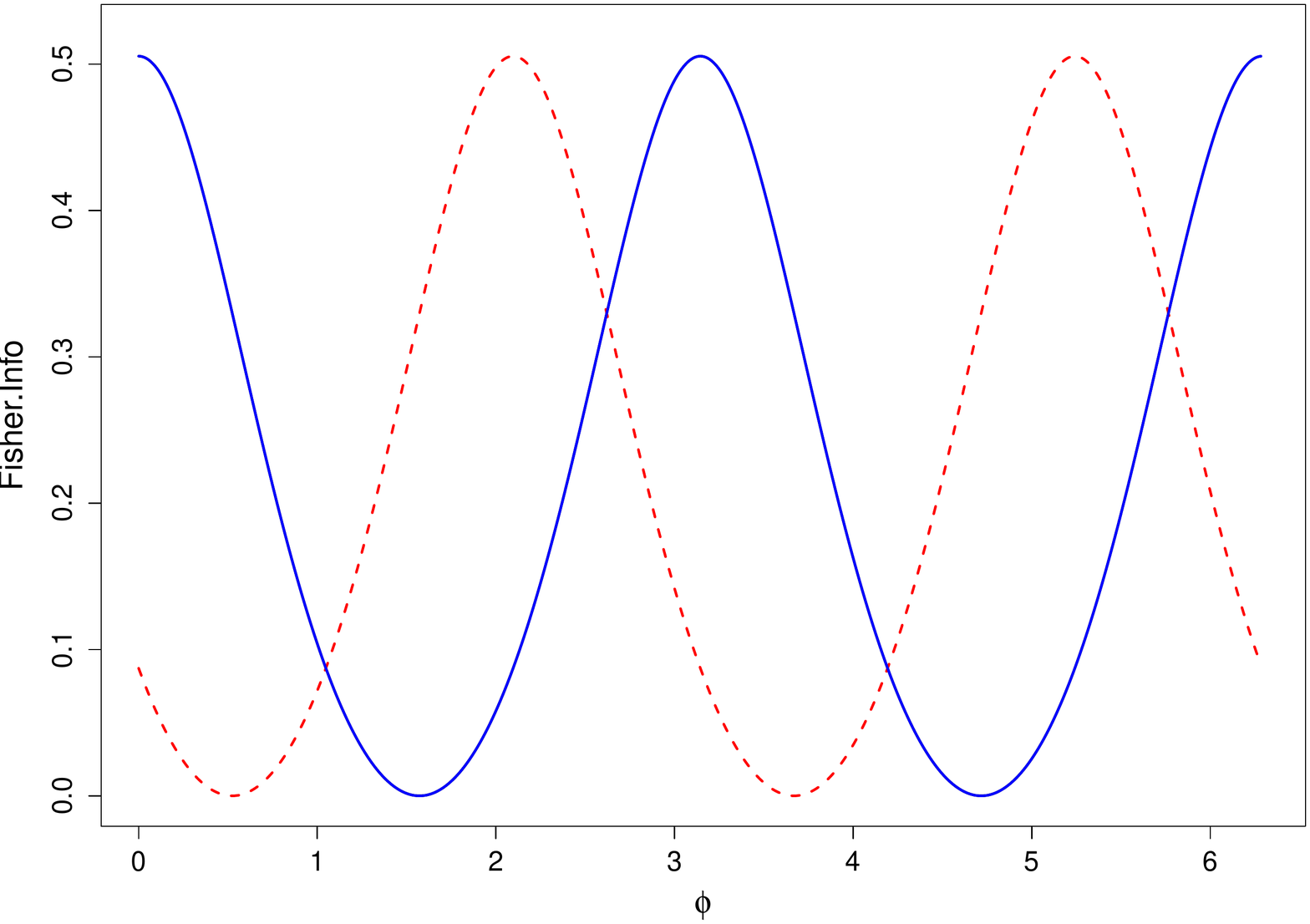}
    \end{center}
  \end{minipage}
  \hfill
  \begin{minipage}[t]{.45\textwidth}
    \begin{center}  
\includegraphics[width=0.99\textwidth]{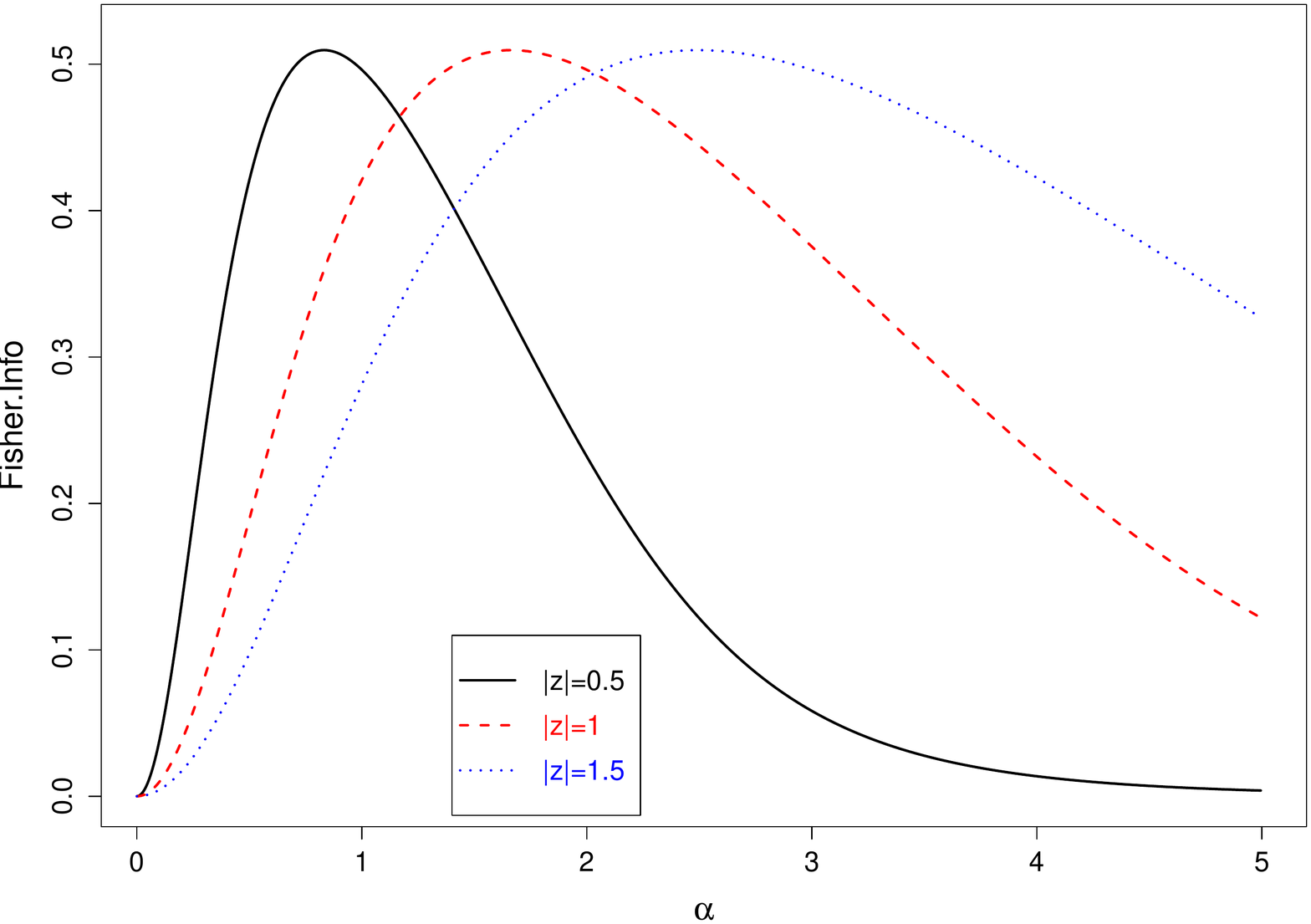}
    \end{center}
  \end{minipage}
  \hfill
\caption{Fisher Information for homodyne measurements : left - dependence on $\phi$ for $|z|/\theta_0 =0.66$ with $\theta=0$ (blue line) and $\theta=\pi /3$  (red dotted line) ; right - dependence on $\theta_0$ for $\theta=\phi=0$ and different laser strengths  }
\end{figure}

In Figure ~\ref{fig:homd} we plot the Fisher information for homodyne detection  as a function of $\phi$ for all other parameters fixed (left)  and as a function of 
$\theta_0$ for fixed $|z|$, $\theta$ and $\phi$ (right). The Fisher information is a function of the ratio $|z|/\theta_0$ and of $cos^2 (\phi+\theta_0)$ therefore it is maximized whenever the laser and the detector are aligned in such a way that $cos(\phi+\theta_0)=\pm1$.

\subsection{The atom maser}\label{sec:atommsr}

A one-atom maser  \cite{Eng02} consists of a beam of excited two level atoms interacting resonantly with a single mode of an electromagnetic field enclosed in a dissipative cavity. The `system' is the field in the cavity, and in a certain time coarse graining approximation, the interaction with a Poissonian beam of atoms, and that with a positive temperature thermal bath can be modelled as a coupling to 4 bosonic channels, one for each possible interaction. Two channels correspond to the two possible outcomes of the excited atom passing through the cavity and the other two channels are the photon exchange channels between the cavity and a thermal bath of constant temperature. The Lindblad generator is
\begin{equation}
\mathcal{L}(X) = \sum_{i=1}^{4}
L^{*}_{i,\phi} X  L_{i,\phi} - \frac{1}{2} \{ L_{i,\phi}^{*} L_{i,\phi} ,X \}  
\end{equation}
with the  four jump operators defined as
\begin{align} L_{1,\phi} &= \sqrt{N_{ex}} a^{*} \frac{\sin(\phi \sqrt{a a^{*}})}{\sqrt{a a^{*}}}, \quad 
L_{2,\phi} = \sqrt{N_{ex}} \cos(\phi \sqrt{a a^{*}}),\\
		L_{3,\phi} &= \sqrt{\nu+1}a, \quad L_{4,\phi} = \sqrt{\nu} a^{*},
\end{align}
where $N_{ex}$ is the rate of the incoming atoms, $\nu$ is the average number of photons in the bath, $a$ and $a^*$ creation and annihilation operators for the field and $\phi$ is the accumulated Rabi angle which is proportional to the interaction strength. 

The atom maser has a unique stationary state which is diagonal in the number basis, and its coefficients are
\begin{equation}
		\rho_{ss}(n)=\rho_{ss}(0)\prod_{i=1}^{n}{\left[\frac{\nu}{\nu+1}+\frac{N_{ex}}{\nu+1}\frac{\sin^{2}{\phi \sqrt{i}}} {i}\right]}.
\end{equation}
The large deviations theory and the central limit theorem for counting measurements has been studied in \cite{vanHorssenGuta}, while the problem of estimating the parameter $\phi$ has been investigated in detail in \cite{CatanaGutavanHorssen12} and \cite{CatanaGutaKypraios}.

\begin{figure}\label{fig:fisherinfo}
  \hfill
  \begin{minipage}[t]{.45\textwidth}
    \begin{center}  
       \includegraphics[width=0.99\textwidth]{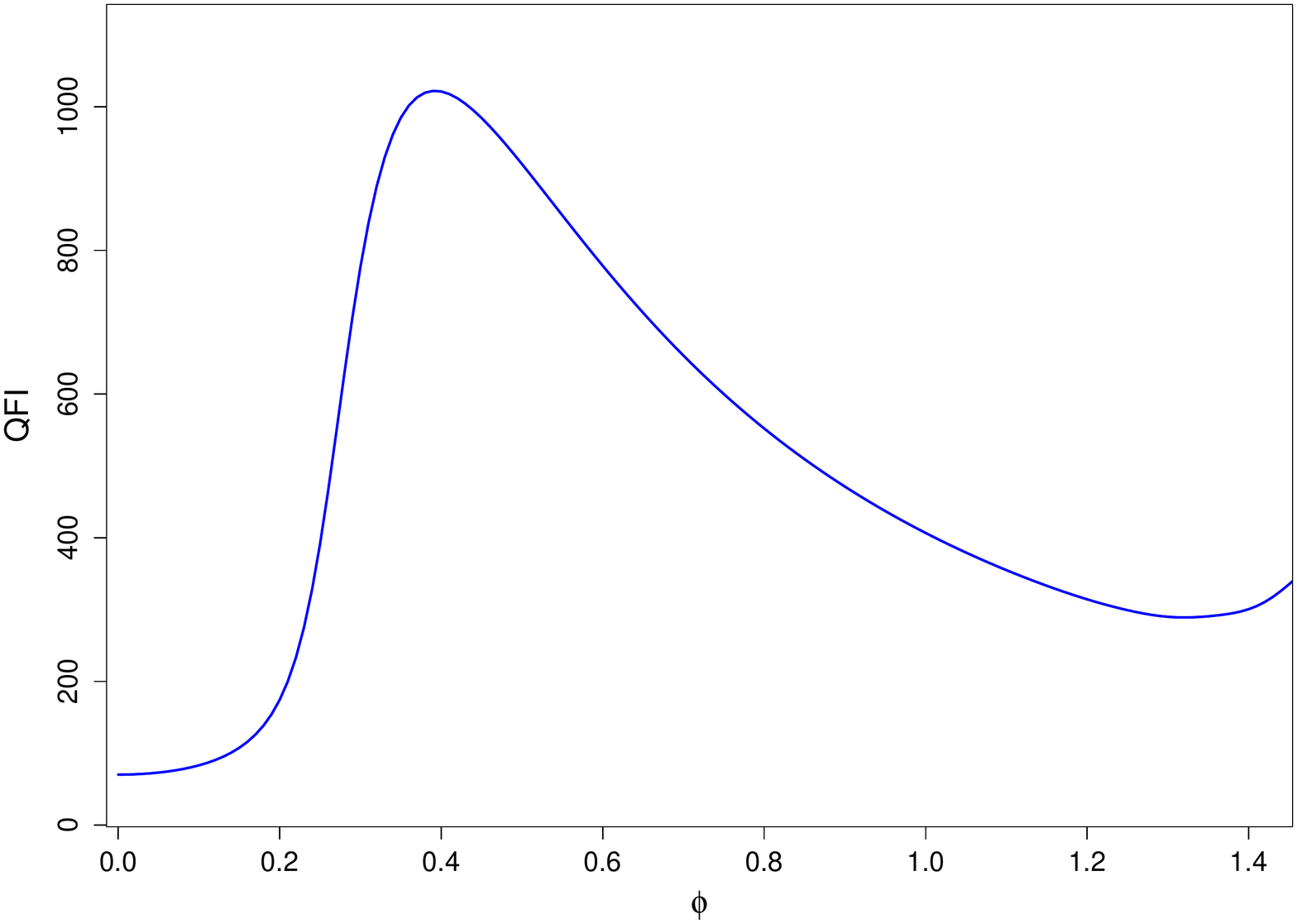}
    \end{center}
  \end{minipage}
  \hfill
  \begin{minipage}[t]{.45\textwidth}
    \begin{center}  
\includegraphics[width=0.99\textwidth]{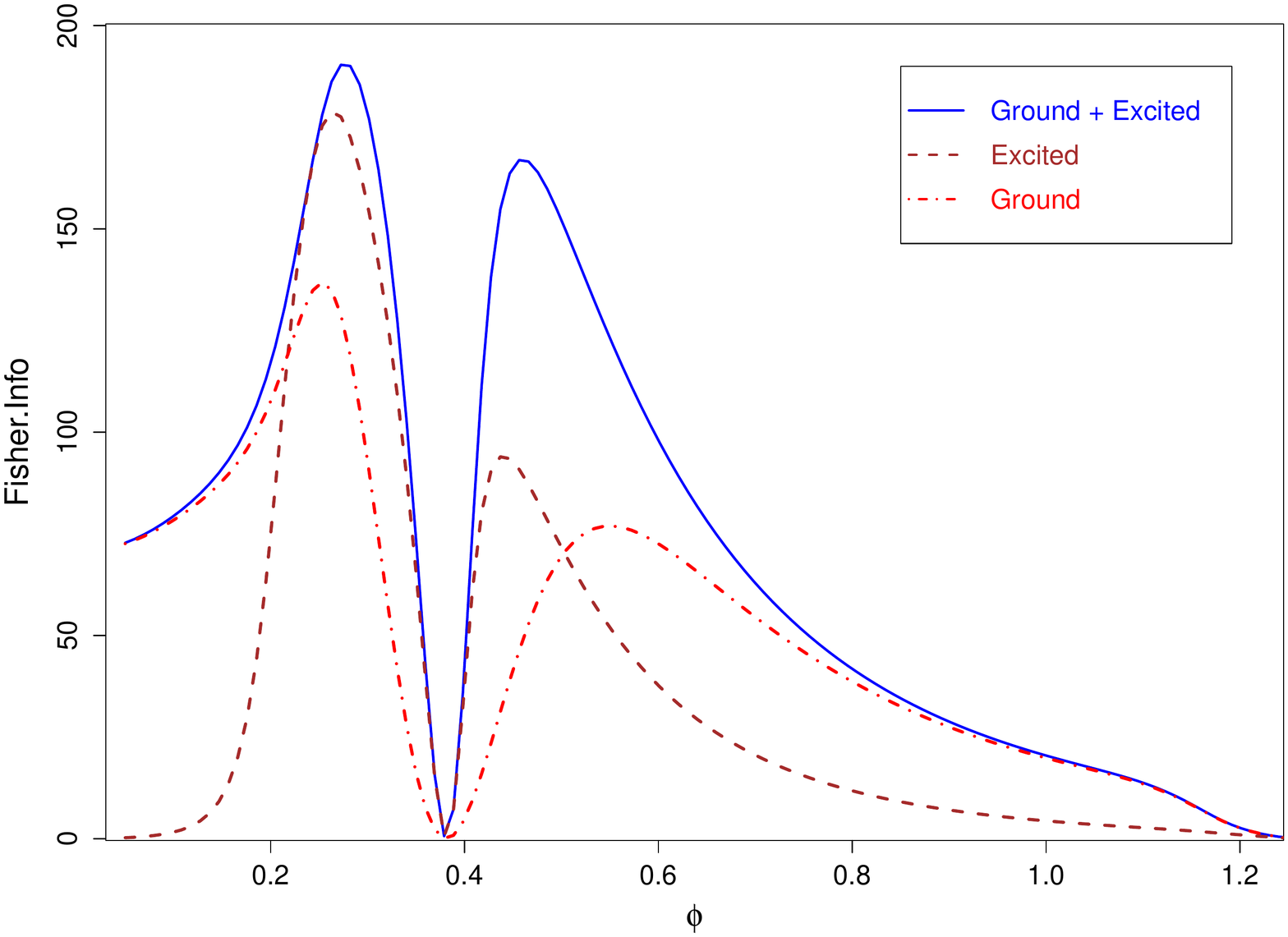}
    \end{center}
  \end{minipage}
  \hfill
\caption{The quantum Fisher information of the atom maser as function of the Rabi angle $\phi$ (left panel). Classical Fisher information corresponding to the total counts of ground state atoms, excited state atoms, and joint statistics (right panel).}
\label{fig:counts}
\end{figure}

In \cite{CatanaGutavanHorssen12} it was shown that the quantum Fisher information described in Theorem \ref{qlan} is
\begin{equation}
F=4 \rm Tr \left( \rho_{ss} \sum_{i=1} ^4 \dot{L}_i^{*}\dot{L}_i \right)= 
4 N_{ex}\rm Tr\left( \rho_{ss} a a^{\dagger}\right)=4 N_{ex}\sum_k (k+1)  \rho_{ss}(k).
\end{equation}
This is plotted in the left panel of figure ~\ref{fig:counts} for $N_{ex}=16$ and $\nu=0.1$. For comparison, the classical Fisher informations associated to total counts of ground state atoms, excited state atoms and ground and excited state jointly, is plotted in the right panel. We note that all informations are equal to zero at a particular value $\phi_0$ where 
the mean photon number in the cavity, and the rate of  ground state atoms are at a maximum, and therefore the derivative with respect to $\phi$ is zero. One can show however \cite{CatanaGutaKypraios}, that the Fisher information of the \emph{full detection record} is strictly larger than zero for all $\phi$. This shows the importance of extending the theory developed here for total counts and time integrated statistics, to more general statistics depending on time correlations.

%


Besides counting, one could in theory consider that a homodyne measurement is performed on the photon loss channel. However this is equal to zero as the homodyne current has mean zero. Interestingly, if the homodyne measurement is performed jointly with the counting measurement on the atomic channels, then the joint classical Fisher information is slightly larger than that of the counting measurement alone. This is due to small non-diagonal terms in the covariance matrix of the two statistics.

\section{Conclusion}

In this paper we have extended the discrete-time results of  \cite{Guta11, Guta&Kiukas14} to the domain of continuous-time quantum Markov processes, in the input-output formalism. For an irreducible system whose dynamics depends on an unknown parameter, we have have shown that for large time, the output state can be approximate by a quantum Gaussian state (asymptotic normality) and found the explicit expression of the asymptotic quantum Fisher information of this state. This provides an absolute bound on the estimation precision of any measurement procedure.
 
We have then analysed the statistical properties of the counting and homodyne continuous-time  measurements. We showed that the total counts and the integrated homodyne current also satisfy asymptotic normality and computed the general expression of the corresponding classical Fisher informations. We then considered two examples (two level system and the atom maser), in which the performance of these measurements is compared with that prescribed by the quantum Fisher information. Finding the optimal measurement and estimation scheme is an important open problem which goes beyond the scope of this paper. Another remaining problem is to extend the present results to a multi-dimensional parameter set-up, and to derive the general quantum Central Limit Theorem which underpins the asymptotic normality results \cite{GutaKiukas2}.

\emph{Acknowledgements.} This work was supported by the EPSRC grant EP/J009776/1.
\section{Appendix}
\subsection{Details of the proof of Theorem \ref{qlan}}\label{proof.qlan}

We apply Theorem \ref{th.semigroup.limit} for the family of semigroups $T_\tau^{(t,u,v)}$ where $t$ plays the role of 
index (instead of the discrete index $n$).
The generator in Eq. \eqref{eq:qgen} can be expanded as 

\begin{equation}\label{eq.generator.qlan}
 \mathcal{L}^{(t,\theta,\theta^\prime)}(X) = t \mathcal{L}_0(X) +
 \sqrt{t} \mathcal{L}_1^{(u,v)}(X) + \mathcal{L}^{(u,v)}_2(X)  + O(t^{-1/2})
\end{equation}
we get 
  \begin{eqnarray*}
\mathcal{L}_0(X) & = &   i[H ,X] +\sum_{i=1}^d \left(  L_{i}^* X L_{i,} - 
   \frac{1}{2}\{L_{i}^* L_{i},X \}\right),\\
\mathcal{L}_1(X) & = &   i(u\dot{H} X - v X \dot{H} )+ \sum_{i=1}^d
 \left( u \dot{L}_i^{*}X L_i + v L_i^{*}X \dot{L}_i - \frac{1}{2}\left(u(\dot{L}_i^{*}L_i+L_i^{*}\dot{L}_i)X+
 vX(\dot{L}_i^{*}L_i+L_i^{*}\dot{L}_i)\right ) \right)\\
&& -i(u-v) \left\langle  \dot{H}+{\rm Im} \sum_{i=1}^{d}\dot{L}_i^{*}L_i\right\rangle_{ss} X  ,\\
\mathcal{L}_2(X) &=&\frac{i}{2}(u^2\ddot{H} X - v^2X\ddot{H} ) +
\sum_{i=1}^d \left(  \frac{u^2}{2} \ddot{L}_i^{*}X L_i+\frac{v^2}{2} L_i^{*}X \ddot{L}_i+uv \dot{L}_i^{*}X \dot{L}_i\right)\\
&&-\frac{1}{2}\sum_{i=1}^d \left\{  \frac{u^2}{2} (\ddot{L}_i^{*} L_i+L_i^{*} \ddot{L}_i +2\dot{L}_i^{*} \dot{L}_i)X +
\frac{v^2}{2}X (\ddot{L}_i^{*} L_i+L_i^{*} \ddot{L}_i+2\dot{L}_i^{*} \dot{L}_i)\right\}. 
\end{eqnarray*}

Since ${\rm Ker}(\mathcal{L}_0)= \mathbb{C} \mathbf{1}$  we choose the projection 
$P(X)=Tr(\rho_{ss}X) \mathbf{1}$ where $\rho_{ss}$ is the stationary state of the system. 
Then, since $P$ is a one dimensional projection, the limit generator in Theorem \ref{th.semigroup.limit} is of the form 
\begin{equation}\label{eq.generator.limit}
\mathcal{L}^{(u,v)}:= -P\mathcal{L}_1\tilde{\mathcal{L}}\mathcal{L}_1 +P\mathcal{L}_2P = 
f(u,v)P
\end{equation}
where $f(u,v)$ is a (complex valued) function. The limit of the inner product in \eqref{eq.inner.products} is therefore 
given by
$$
\lim_{t\to \infty}\Braket { \tilde{\Psi}_t^u |{ \tilde{\Psi}_t^v }}
=
\lim_{t\to \infty} \left\langle{\chi_0}\left| T_{1}^{(t,u,v)}(\mathbf{1}) \right|{\chi_0}\right\rangle=
\left\langle{\chi_0}\left| T_{1}^{(u,v)}(\mathbf{1}) \right|{\chi_0}\right\rangle = e^{f(u,v)}.
$$
We will now calculate the function $f(u,v)$. For the second term in \eqref{eq.generator.limit} we have
\begin{equation*}\begin{split}
&\mathcal{L}_2(\mathbf{1})=(u^2-v^2) \left(\frac{i}{2} \ddot{H}+ \frac{1}{4}\sum_i(\ddot{L}_i^{*}L_i-L_i^*\ddot{L}_i) \right)- \frac{1}{2}(u-v)^2 \sum_i \dot{L}_i^{*} \dot{L}_i\\
&= \frac{i}{2}(u^2-v^2)\left(\ddot{H}+ {\rm Im} \sum_i \ddot{L}_i^{*}L_i\right)- \frac{1}{2}(u-v)^2 \sum_i \dot{L}_i^{*} \dot{L}_i
\end{split}\end{equation*}

  \begin{equation*}\begin{split}
\mathcal{L}_1(\mathbf{1})&= (u-v)(i\dot{H}+\frac{1}{2}\sum_i (\dot{L}_i^{*}L_i- L_i^*\dot{L}_i)
-i(u-v) 
\left\langle \dot{H}+{\rm Im} \sum_{i=1}^{n}\dot{L}_i^{*}L_i\right\rangle_{ss} \\
&=i(u-v)\left[\dot{H}+{\rm Im}\sum_i \dot{L}_i^{*}L_i- 
 \left\langle \dot{H}+{\rm Im}
\sum_{i=1}^{n}\dot{L}_i^{*}L_i)\right\rangle_{ss}\right]
\equiv i(u-v)B 
  \end{split}\end{equation*}
Note that the condition $\rm Tr(\mathcal L _1(\mathbf1) \rho_{ss})=0$ is satisfied. 
For simplicity we denote $\tilde{B} :=  \tilde{\mathcal{L} }(B)$, and note that $\langle \tilde{B}\rangle_{ss}=0$ since 
$\tilde{\mathcal{L}}$ leaves the set of zero expectation observables invariant.
Rearranging the terms in a suitable way we find that 
\begin{equation*}\begin{split}
\mathcal{L}_1  \tilde{\mathcal{L}} \mathcal{L}_1 (\mathbf{1})
=& i(u^2-v^2)\left \{ -{\rm Im}(\dot{H} \tilde{B})+ {\rm Re}(\sum_i \dot{L}_i^{*}\tilde{B}L_i) - {\rm Re}(\sum_i \dot{L}_i^{*}L_i)\tilde{B} \right \}-\\
& -(u-v)^2 \left \{  {\rm Re}(\dot{H}\tilde{B})+ {\rm Im}(\sum_i \dot{L}_i^{*}\tilde{B}L_i) 
- \left\langle \dot{H}+{\rm Im} \sum_{i=1}^{n}\dot{L}_i^{*}L_i\right\rangle_{ss} \tilde{B} \right \} .
\end{split}\end{equation*}
Therefore we find 
$$
f(u,v)=
\exp\left( i(u^2-v^2) \langle X_2 \rangle_{ss} - (u-v)^2  \langle X_1 \rangle_{ss} \right),
$$
where $X_1$ and $X_2$ are the selfadjoint operators
\begin{eqnarray*}
X_1&=& \frac{1}{2}\sum_i \dot{L}_i^{*}\dot{L}_i - {\rm Re}(\dot{H} \tilde B)- {\rm Im}(\sum_i \dot{L}_i^{*}\tilde B L_i),\\
X_2&=& \frac{1}{2}(\ddot{H}+ {\rm Im}\sum_i \ddot{L}_i^{*}L_i)  +{\rm Im}(\dot{H}\tilde B)- {\rm Re}(\sum_i \dot{L}_i^{*}\tilde B L_i)+ {\rm Re}(\sum_i \dot{L}_i^{*}L_i)\tilde B.
\end{eqnarray*}
We conclude that the limit overlap can be expressed in terms of the overlap of two one-mode coherent states, with a certain choice of phase which does not have a physical significance
$$
\lim_{t\to \infty}\Braket { \tilde{\Psi}_t^u |{ \tilde{\Psi}_t^v }} = 
e^{-(u-v)^2 F/8} e^{ i(u^2-v^2) \langle X_2 \rangle_{ss}} =
e^{ i(u^2-v^2) \langle X_2 \rangle_{ss}}
\braket{\sqrt {F/2} u |\sqrt{F/2} v}.
$$
The constant $F= 8 \rm \langle X_1\rangle_{ss}$ is the quantum Fisher information of the limiting coherent state model 
$\{ |\sqrt{F/2} u\rangle \,: \, u\in \mathbb{R} \}$. This completes the proof of the quantum local asymptotic normality Theorem \ref{qlan}.

\subsection{Details of the proof of Theorem \ref{lancounts}}\label{app.counting}
In the following we drop the subscript when $\theta= \theta_0$ and denote $L= L_{\theta_0}, H=H_{\theta_0} $. By differentiating \eqref{eq.semigroup.rescaled.counting} and using the quantum Ito rules it can be checked that $S_\tau^{(t,u,s)}$ is a semigroup with generator
\begin{eqnarray*}
\mathcal{L}^{(t,u,s)} (X)&=& 
t\left[ \mathcal{L}_{\theta } (X) + (e^{is/\sqrt{t}}-1 ) L_\theta^* X L_\theta - \frac{is}{\sqrt{t}}\langle L^{*}L\rangle_{ss}  X \right] \\
&=&
 t \mathcal{L}_0 (X)+ \sqrt{t} \mathcal{L}_1^{(u,s)}(X) + \mathcal{L}^{(u,s)}_2(X)  + O(t^{-1/2}).
\end{eqnarray*}
where
  \begin{eqnarray*}
\mathcal{L}_0(X) &=&   i[H,X] +  L^{*}X L - \frac{1}{2}\{L^{*}L,X\} ,\\
\mathcal{L}_1(X) &=&  iu[\dot{H},X]+  is L^{*}XL + u ( \dot{L}^{*}XL+ L^{*}X\dot{L})\\
 & &-\frac{u}{2}\left( (  \dot{L}^{*}L+ L^{*}\dot{L} )X + X( \dot{L}^{*}L + L^{*}\dot{L}  )\right) -is\langle L^{*}L\rangle_{ss}  X
 \\[1mm]
 \mathcal{L}_2(X) &=&\frac{iu^2}{2}[\ddot{H} , X] - 
  \frac{s^2}{2} L^{*}X L  + isu( \dot{L}^{*}XL + L^{*}X\dot{L})+\\
&& +\frac{u^2}{2} (\ddot{L}^{*}XL + L^{*}X\ddot{L}+2 \dot{L}^{*}X\dot{L} ) - \\
&& -\frac{u^2}{4} (\ddot{L}^{*}L X + L^{*}\ddot{L}X+2\dot{L}^{*}\dot{L}X+X\ddot{L}^{*}L + XL^{*}\ddot{L} + 2X\dot{L}^{*}\dot{L}). 
\end{eqnarray*}
Therefore we have
\begin{equation}\label{eq:l11}
\mathcal{L}_1(\mathbf{1})=is\left( L^{*}L -  \langle L^*L \rangle_{ss} \mathbf{1}\right), \quad 
\mathcal{L}_2(\mathbf{1})=-\frac{s^2}{2}L^{*}L + isu (\dot{L}^{*} L + L^{*}\dot{L}).
\end{equation}
and in particular ${\rm Tr}(\rho_{ss} \mathcal{L}_1(\mathbf{1})) =0$. Since the dynamics is irreducible at $\theta_0$, 
we have ${\rm Ker}(\mathcal{L}_0)= \mathbb{C} \mathbf{1}$ and the conditions of Theorem \ref{th.semigroup.limit} are fulfilled with $P$ being the projection $P(X)=Tr(\rho_{ss}X) \mathbf{1}$. Therefore 
$$
\lim_{t\to \infty} S_\tau^{(t,u,s)}(\mathbf{1}) = S_\tau^{(u,s)} (\mathbf{1}) =  \exp(\tau f_c(s,u) ) \mathbf{1}
$$
where $S_\tau^{(u,s)}$ is a semigroup on the one dimensional space $ \mathbb{C} \mathbf{1}$  with generator 
$$
\mathcal{L}^{(u,s)}:= -P\mathcal{L}_1\tilde{\mathcal{L}}\mathcal{L}_1 +P\mathcal{L}_2P = f_c(s,u) P
$$ 
It now remains to compute the function $ f_c(s,u) $. 
With the above expressions for $\mathcal{L}_1(\mathbf{1})$ and $\mathcal{L}_2(\mathbf{1})$ we get 
$$
\left[-P\mathcal{L}_1\tilde{\mathcal{L}}\mathcal{L}_1 +  P\mathcal{L}_2P \right](\mathbf{1}) = 
(ius \mu_c- \frac{s^2}{2} V_c)\mathbf{1}
$$
where 
\begin{eqnarray*}
\mu_c &:=& \left\langle i[\dot{H},A]+  \dot{L}^{*}AL + L^{*}A\dot{L} 
+ 2 {\rm Re}( \dot{L}^{*}L) 
- ({\rm Re}( \dot{L}^{*}L) A + A {\rm Re}( \dot{L}^{*}L))\right\rangle_{ss},\\
V_c &:=&\left\langle L^{*}L + 2 L^{*} A L  
\right\rangle_{ss}.
\end{eqnarray*}
with $A$ given by 
$A := -\tilde{\mathcal{L}} \left(  L^{*}L -\langle L^*L \rangle_{ss}\right)$. In the expression of final expression of 
$V_c$ we used the fact that $\langle A\rangle_{ss}=0$ since $\tilde{\mathcal{L}} $ leaves the space of zero mean observables invariant.

\subsection{Details of the proof of Theorem \ref{lanhom}}\label{app.homo}

The generator of the semigroup $T_\tau^{(t,u,p)}$ is
\begin{eqnarray*}
  \mathcal{L}^{(t,u,p)}(X) &=& 
  t\left[ i[H_\theta,X]+
  L_{\theta}^{*}XL_\theta -
  \frac{1}{2}\left(L_{\theta}^{*}L_\theta X + X L_{\theta}^{*} L_\theta \right)
  \right. \\
&& \left. + \frac{ip}{\sqrt t}(e^{-i\phi}L_{\theta}^{*}X+ Xe^{i\phi}L_{\theta} ) 
- \frac{p^2}{2t}X- \frac{ip}{\sqrt t} \langle e^{-i\phi}L_{\theta_0}^{*}+ e^{i\phi}L_{\theta_0}\rangle_{ss} X\right]\\
&=& t \mathcal{L}_0 (X)+ \sqrt{t} \mathcal{L}_1^{(u,p)}(X) + \mathcal{L}^{(u,p)}_2(X)  + O(t^{-1/2}).
\end{eqnarray*}

In the following we drop the subscript when $\theta= \theta_0$ and denote $L= L_{\theta_0}, H=H_{\theta_0} $. The first three terms of the generator are
  \begin{eqnarray*}
\mathcal{L}_0(X) &=&   i[H,X] +  L^{*}X L - \frac{1}{2}\{L^{*}L,X\} ,\\
\mathcal{L}_1(X) &=&  iu[\dot{H},X]+   u ( \dot{L}^{*} X L + L^{*} X \dot{L})  
-\frac{u}{2}\left( (  \dot{L}^{*}L + L^{*}\dot{L}  )X+X( \dot{L}^{*}L + L^{*}\dot{L} )\right) \\
&& + ip(e^{-i\phi}L^{*}X + e^{i\phi}X L ) -i p\langle e^{-i\phi}L^{*}+ e^{i\phi}L \rangle_{ss} X \\
\mathcal{L}_2(X) &=& \frac{iu^2}{2}[\ddot{H},X] + iup(e^{-i\phi}\dot{L}^{*}X+ e^{i\phi}X\dot{L}  ) 
+\frac{u^2}{2} (\ddot{L}^{*}XL+L^{*}X\ddot{L}+2 \dot{L}^{*}X\dot{L})  \\
&& -\frac{u^2}{4} (\ddot{L}^{*}L X+L^{*}\ddot{L} X+2\dot{L}^{*}\dot{L} X+X \ddot{L}^{*}L+XL^{*}\ddot{L}+2X\dot{L}^{*}\dot{L})- \frac{p^2}{2}X  . 
\end{eqnarray*}
Then
\begin{eqnarray*}
\mathcal{L}_1(\mathbf{1})&=& ip(e^{-i\phi}L^{*}+ e^{i\phi}L) -i p\langle e^{-i\phi}L^{*}+ e^{i\phi}L\rangle_{ss}\mathbf{1} ,\\
\mathcal{L}_2(\mathbf{1})&=& iup(e^{-i\phi}\dot{L}^{*}+ e^{i\phi}\dot{L})-\frac{p^2}{2}\mathbf{1}.
\end{eqnarray*}
and in particular ${\rm Tr}(\rho_{ss} \mathcal{L}_1(\mathbf{1})) =0$. Since the dynamics is irreducible, we have 
${\rm Ker}(\mathcal{L}_0)= \mathbb{C} \mathbf{1}$ and the conditions of Theorem \ref{th.semigroup.limit} are fulfilled with $P$ being the projection $P(X)=Tr(\rho_{ss}X) \mathbf{1}$. Therefore 
$$
\lim_{t\to \infty} T_\tau^{(t,u,p)}(\mathbf{1}) = T_\tau^{(u,p)} (\mathbf{1}) =  \exp(\tau f_h(u,p) ) \mathbf{1}
$$
where $T_\tau^{(u,p)}$ is a semigroup on the one dimensional space $ \mathbb{C} \mathbf{1}$  with generator 
$$
\mathcal{L}^{(u,p)}:= -P\mathcal{L}_1\tilde{\mathcal{L}}\mathcal{L}_1 +P\mathcal{L}_2P = f_h(u,p) P.
$$ 
It now remains to compute the function $ f_h(u,p) $. 
With the above expressions for $\mathcal{L}_1(\mathbf{1})$ and $\mathcal{L}_2(\mathbf{1})$ and with the notation 
$ipB \equiv \tilde{\mathcal{L}}\mathcal{L}_1(\mathbf{1})$ such that ${\rm Tr}(\rho_{ss}B)=0$, we have
\begin{equation*}
e^{ -P\mathcal{L}_1\tilde{\mathcal{L}}\mathcal{L}_1 +  P\mathcal{L}_2 P }(\mathbf{1})= e^{iup \mu_h- \frac{p^2}{2} V_h}
\mathbf{1},
\end{equation*}
where
\begin{eqnarray*}
\mu_h&=&\left\langle 
i[\dot{H} , B]+  \dot{L}^{*}BL+ L^{*}B\dot{L}+ e^{-i\phi}\dot{L}^{*}+ e^{i\phi}\dot{L} 
-\frac{1}{2}\left(( \dot{L}^{*}L + L^{*}\dot{L}  )B+B( \dot{L}^{*}L + L^{*}\dot{L} )\right) \right\rangle_{ss},\\
V_h & =&  1+ 2  \langle e^{-i\phi}L^{*}B + e^{i\phi} BL\rangle_{ss} 
\end{eqnarray*}

\subsection{Calculations for the two level system}\label{app.2levels}

With respect to the basis \eqref{eq.basis}, the map $\mathcal{L}_0$ has the matrix representation

  \begin{equation}
\mathcal{L}_0 = \begin{pmatrix}
  -\theta_0^2      & \bar{z}\theta_0       & z\theta_0             & 0\\
  -2z\theta_0        & -\frac{\theta_0^2}{2} & 0                   & 0\\
  -2\bar{z}\theta_0  &   0                 & -\frac{\theta_0^2}{2} & 0\\
  0                &   0                 & 0                   & 0
  \end{pmatrix}.
  \end{equation}
Its kernel is
$\mathbb{C}\mathbf{1}$ and the projection $P$ is given by $P(X) = \gamma_4 \mathbf{1}= {\rm Tr} (\rho_{ss}X)$, where
$X=\gamma_1 e_1+\gamma_2 e_2+\gamma_3 e_3+\gamma_4 e_4$ is the expansion of $X$ in the basis \eqref{eq.basis}.
From \eqref{eq.els} we get
  \begin{eqnarray*}
  \mathcal{L}_1(\mathbf{1}) &=& (v-u)(z\sigma_+-\bar{z}\sigma_-) = 
  (v-u)(ze_2-\bar{z}e_3).\\
P\mathcal{L}_2 (\mathbf{1}) &=& -\frac{(u-v)^2}{2}P(\sigma_+\sigma_-) = 
  -\frac{(u-v)^2}{2}\frac{1-\tilde{a}}{2}\mathbf{1}.  
  \end{eqnarray*}

where we have used that $\sigma_+\sigma_- = \frac{1}{2}e_1 + \frac{1-\tilde{a}}{2}e_4$. To compute the other term in the Gaussian kernel we need to invert $\mathcal{L}_0$ on the complement of $P$. With respect to the basis $(e_1,e_2,e_3)$ the map $\tilde{\mathcal{L}}$ is given by 
  \begin{equation*}
  \tilde{\mathcal{L}} =\frac{1}{\theta_0^2(\theta_0^2+8|z|^2)}
  \begin{pmatrix}
  -\theta_0^2 & -2\bar{z}\theta_0 & -2z\theta_0\\
  4z\theta_0 & -2\theta_0^2-8|z|^2 & 8 z^2\\
  4\bar{z}\theta_0 & 8 \bar{z}^2 & -2\theta_0^2 - 8|z|^2
  \end{pmatrix}. 
  \end{equation*}

Therefore
  \begin{equation*}
  \tilde{\mathcal{L}}\mathcal{L}_1(\mathbf{1}) = (v-u)\left(
  \frac{16|z|^2\bar{z} + 2\theta_0^2 \bar{z}}{\theta_0^2(\theta_0^2+8|z|^2)} e_3 - 
  \frac{16|z|^2 z + 2\theta_0^2 z}{\theta_0^2(\theta_0^2+8|z|^2)} e_2
  \right),
  \end{equation*}
and then
  \begin{eqnarray*}
  P\mathcal{L}_1\tilde{\mathcal{L}}\mathcal{L}_1(\mathbf{1}) &=& (u-v)
  \frac{16|z|^2 + 2\theta_0^2}{\theta_0^2(\theta_0^2+8|z|^2)}
  P\mathcal{L}_1(ze_2 - \bar{z}e_3) 
  =
  \frac{2}{\theta^2_0}(u-v)P\mathcal{L}_1(ze_2 - \bar{z}e_3) \\
  &=& -\frac{2(8|z|^2 - \theta_0^2)|z|^2}{(8|z|^2 + \theta_0^2)\theta_0^2}(u-v)^2\mathbf{1}.
  \end{eqnarray*}  
Putting everything together we find
  \begin{equation*}\label{eq.cont.time.u-v.2.coefficient}
  -P\mathcal{L}_1\tilde{\mathcal{L}}\mathcal{L}_1(\mathbf{1}) + P\mathcal{L}_2 (\mathbf{1}) =
  -\frac{16|z|^4}{(8|z|^2 + \theta_0^2)\theta_0^2}(u-v)^2 \mathbf{1},
  \end{equation*}
which means that the quantum Fisher information is 
\begin{equation}
F(\theta_0)=\frac{128 |z|^4}{(8|z|^2 + \theta_0^2)\theta_0^2}. 
\end{equation}

%

\begin{thebibliography}{10}

\bibitem{Dowling&Milburn}
J.~P. Dowling and G.~J. Milburn.
\newblock {Quantum technology: the second quantum revolution}.
\newblock {\em Phil. Trans. R. Soc. Lond. A}, 361:1655--1674, 2003.

\bibitem{HarocheRaimond}
P.~Haroche and J.-M. Raimond.
\newblock {\em {Exploring the Quantum: Atoms, Cavities, and Photons}}.
\newblock Oxford University Press, 2006.

\bibitem{Brassard01}
G~Brassard.
\newblock Quantum communication complexity (a survey).
\newblock In A.~Gonis and P.E.A. Turchi, editors, {\em Decoherence and its
  implications in quantum computing and information transfer. Proc Nato
  advanced research Workshop, Mykonos, Greece 25-30 June 2000}, pages 199--210.
  IOS Press, Amsterdam, 2001.

\bibitem{NC00}
M.A. Nielsen and I.L. Chuang.
\newblock {\em Quantum Computation and Quantum Information}.
\newblock Cambridge University Press, Cambridge, 2000.

\bibitem{Giovannetti&Science04}
V.~Giovannetti, S.~Lloyd, and L.~Maccone.
\newblock Quantum-enhanced measurements: beating the standard quantum limit.
\newblock {\em Science}, 306:1330, 2004.

\bibitem{Gardiner&Zoller}
C.~Gardiner and P.~Zoller.
\newblock {\em {Quantum Noise}}.
\newblock Springer, 2004.

\bibitem{Car93}
H.~J. Carmichael.
\newblock {\em An Open Systems Approach to Quantum Optics}.
\newblock Springer-Verlag, Berlin Heidelberg New-York, 1993.

\bibitem{Mabuchi&Khaneja}
H.~Mabuchi and N.~Khaneja.
\newblock {Principles and applications of control in quantum systems}.
\newblock {\em Int. J. Robust Nonlinear Control}, 15:647--667, 2005.

\bibitem{Wiseman&Milburn}
H.~M. Wiseman and G.~J. Milburn.
\newblock {\em {Quantum Measurement and Control}}.
\newblock Cambridge University Press, 2009.

\bibitem{Breuer02}
H-P. Breuer and F.~Petruccione.
\newblock {\em The theory of open quantum systems}.
\newblock Oxford University Pressl, 2002.

\bibitem{Mab96}
H.~Mabuchi.
\newblock Dynamical identification of open quantum systems.
\newblock {\em Quantum Semiclass. Opt.}, 8:1103--1108, 1996.

\bibitem{Wiseman&Gambetta}
J.~Gambetta and H.M. Wiseman.
\newblock State and dynamical parameter estimation for open quantum systems.
\newblock {\em Phys.Rev. A}, 64:042105, 2001.

\bibitem{Molmer&Gammelmark}
S.~Gammelmark and K.~Molmer.
\newblock {Bayesian parameter inference from continuously monitored quantum
  systems}.
\newblock {\em Phys. Rev. A}, 87:032115, 2013.

\bibitem{Fujiwara}
A.~Fujiwara.
\newblock {Quantum channel identification problem}.
\newblock {\em Phys. Rev. A}, 63:042304, 2001.

\bibitem{Burgarth}
D.~Burgarth, K.~Maruyama, and F.~Nori.
\newblock {Indirect quantum tomography of quadratic Hamiltonians}.
\newblock {\em New J. of Phys.}, 13:013019, 2011.

\bibitem{Cole}
J.~Cole, S.~Schirmer, A.~Greentree, C.~Wellard, D.~Oi, and L.~Hollenberg.
\newblock {Identifying an experimental two-state Hamiltonian to arbitrary
  accuracy}.
\newblock {\em Phys. Rev. A}, 71:062312, 2005.

\bibitem{Howard}
M.~Howard, J.~Twamley, C.~Wittmann, T.~Gaebel, F.~Jelezko, and J.~Wrachtrup.
\newblock {Quantum process tomography and Linblad estimation of a solid-state
  qubit}.
\newblock {\em New J. of Phys.}, 8:33, 2006.

\bibitem{CatanaGutavanHorssen12}
C.~Catana, M.~van Horssen, and M.~Gu\c{t}\u{a}.
\newblock {Asymptotic inference in system identification for the atom maser}.
\newblock {\em Phil. Trans. R. Soc. Lond. A}, 370:5308--5323, 2012.

\bibitem{CatanaGutaKypraios}
C.~Catana, M.~Gu\c{t}\u{a}, and T.~Kypraios.
\newblock {Maximum likelihood versus likelihood-free quantum system
  identification in the atom maser}.
\newblock {\em J. Phys. A: Math. Theor.}, 47:415302, 2014.

\bibitem{Guta&Kahn}
M.~Gu\c{t}\u{a} and J.~Kahn.
\newblock Local asymptotic normality for qubit states.
\newblock {\em Phys. Rev. A}, 73:052108, 2006.

\bibitem{Guta&Janssen08}
M.~Gu\c{t}\u{a}, B.~Janssens, and J.~Kahn.
\newblock Optimal estimation of qubit states with continuous time measurements.
\newblock {\em Commun. Math. Phys.}, 277:127--160, 2008.

\bibitem{KahnGuta}
J.~Kahn and M.~Gu\c{t}\u{a}.
\newblock {Local asymptotic normality for finite dimensional quantum systems}.
\newblock {\em Commun. Math. Phys.}, 289:597--652, 2009.

\bibitem{Guta&Jencova}
M.~Gu\c{t}\u{a} and A.~Jen\c{c}ov\'{a}.
\newblock {Local asymptotic normality in quantum statistics}.
\newblock {\em Commun. Math. Phys.}, 276:341--379, 2007.

\bibitem{Guta11}
M.~Gu\c{t}\u{a}.
\newblock Quantum fisher information and asymptotic normality in system
  identification for quantum markov chains.
\newblock {\em Phys. Rev . A}, 83:062624, 2011.

\bibitem{Guta&Kiukas14}
M.~Gu\c{t}\u{a} and J.~Kiukas.
\newblock Equivalence classes and local asymptotic normality in system
  identification for quantum markov chains.
\newblock arXiv:1402.3535; to appear in Commun. Math. Phys, 2014.

\bibitem{Hopfner88}
R.~H{\"{o}}pfner.
\newblock Asymptotic inference for continuous-time markov chains.
\newblock {\em Probability Theory and Related Fields}, 77:537--550, 1988.

\bibitem{HuP84}
R.~L. Hudson and K.~R. Parthasarathy.
\newblock Quantum {It\^o's} formula and stochastic evolutions.
\newblock {\em Commun. Math. Phys.}, 93:301--323, 1984.

\bibitem{Par92}
K.~R. Parthasarathy.
\newblock {\em An Introduction to Quantum Stochastic Calculus}.
\newblock Birkh\"auser, Basel, 1992.

\bibitem{Dav80}
E.~B. Davies.
\newblock {\em One-parameter semigroups}.
\newblock Academic Press, London New-York San Francisco, 1980.

\bibitem{Rivas12}
A.~Rivas and S.~Huelga.
\newblock {\em Open quantum system : an introduction}.
\newblock Springer, Berlin, 2012.

\bibitem{Dav76}
E.~B. Davies.
\newblock {\em Quantum Theory of Open Systems}.
\newblock Academic Press, London New-York San Francisco, 1976.

\bibitem{Braunstein&Caves94}
S.L. Braunstein and C.M. Caves.
\newblock Statistical distance and the geometry of quantum states.
\newblock {\em Phys. Rev. Lett.}, 72:3439--3443, 1994.

\bibitem{Molmer}
S.~Gammelmark and K.~M\o{}lmer.
\newblock Fisher information and the quantum cram\'er-rao sensitivity limit of
  continuous measurements.
\newblock {\em Phys. Rev. Lett.}, 112:170401, 2014.

\bibitem{Gill&Guta}
R.~D. Gill and M.~Gu{\c t}{\u a}.
\newblock {On Asymptotic Quantum Statistical Inference}.
\newblock {\em Institute of Mathematical Statistics Collections}, 9:105--127,
  2012.

\bibitem{vanHorssenGuta2}
M.~van Horssen and M.~Gu\c{t}\u{a}.
\newblock {Sanov and Central Limit Theorems for output statistics of quantum
  Markov chains}.
\newblock arXiv:1407.5082, 2014.

\bibitem{Eng02}
B-G. Englert and G.~Morigi.
\newblock Lectures on dissipative master equations.
\newblock In A.~Buchleitner and K.~Hornberger, editors, {\em Coherent evolution
  in noisy environments}, volume 611 of {\em Lecture Notes in Physics}, pages
  55--106. Springer-Verlag, Berlin Heidelberg, 2002.

\bibitem{vanHorssenGuta}
M.~van Horssen and M.~Gu\c{t}\u{a}.
\newblock {Large Deviations, Central Limit and dynamical phase transitions in
  the atom maser}.
\newblock arXiv:1206.4956v2, 2012.

\bibitem{GutaKiukas2}
M.~Gu\c{t}\u{a} and J.~Kiukas.
\newblock in preparation, 2014.

\end{thebibliography}

\end{document}